\documentclass[lettersize,journal,twoside]{IEEEtran}

\PassOptionsToPackage{hyphens}{url}
\usepackage{amsmath,amssymb,amsfonts}
\usepackage{algorithmic}
\usepackage{algorithm}
\usepackage{array}
\usepackage{caption}
\usepackage{textcomp}
\usepackage{stfloats}
\usepackage{url}
\usepackage{verbatim}
\usepackage{graphicx}
\usepackage{cite}
\usepackage{xcolor}
\usepackage[per-mode=symbol,per-symbol =/]{siunitx}
\usepackage[pscoord]{eso-pic}
\usepackage[acronym]{glossaries}
\usepackage{float}
\usepackage{multirow}
\usepackage{setspace}
\usepackage{enumitem}
\usepackage{url}
\usepackage{hhline}
\usepackage[font=footnotesize,labelfont=bf]{caption}
\usepackage[hidelinks]{hyperref}
\usepackage{orcidlink}
\usepackage[capitalise]{cleveref}
\usepackage{placeins}
\usepackage{circledsteps}
\usepackage{tabularx}
\usepackage{threeparttable}
\usepackage{booktabs}
\usepackage{colortbl}
\usepackage{makecell}
\usepackage{subfig}
\usepackage{xspace}
\usepackage{lineno}
\usepackage{tikz}
\usepackage{lipsum}
\usepackage{pifont}
\usepackage[nice]{nicefrac}
\usepackage{arydshln}
\usepackage{soul}
\usepackage{ulem}

\definecolor{ieee-bright-dblue-100}{rgb}{0.0, 0.3828, 0.6055}
\definecolor{ieee-bright-dblue-80}{rgb}{0.0, 0.4883, 0.6797}
\definecolor{ieee-bright-dblue-60}{rgb}{0.3633, 0.6094, 0.7617}
\definecolor{ieee-bright-dblue-40}{rgb}{0.5898, 0.7383, 0.8398}
\definecolor{ieee-bright-dblue-20}{rgb}{0.8906, 0.8984, 0.9219}
\definecolor{ieee-bright-red-100}{rgb}{0.7266, 0.0469, 0.1836}
\definecolor{ieee-bright-red-80}{rgb}{0.832, 0.3164, 0.3281}
\definecolor{ieee-bright-red-60}{rgb}{0.8906, 0.4922, 0.4805}
\definecolor{ieee-bright-red-40}{rgb}{0.9336, 0.6562, 0.6406}
\definecolor{ieee-bright-red-20}{rgb}{0.9688, 0.8203, 0.8125}
\definecolor{ieee-bright-orange-100}{rgb}{0.9961, 0.6367, 0.0}
\definecolor{ieee-bright-orange-80}{rgb}{0.9844, 0.6953, 0.3125}
\definecolor{ieee-bright-orange-60}{rgb}{0.9883, 0.7695, 0.4844}
\definecolor{ieee-bright-orange-40}{rgb}{0.9922, 0.8359, 0.6562}
\definecolor{ieee-bright-orange-20}{rgb}{0.9961, 0.9219, 0.8164}
\definecolor{ieee-bright-yellow-100}{rgb}{0.9961, 0.8164, 0.0}
\definecolor{ieee-bright-yellow-80}{rgb}{0.9961, 0.8477, 0.2148}
\definecolor{ieee-bright-yellow-60}{rgb}{0.9961, 0.875, 0.4492}
\definecolor{ieee-bright-yellow-40}{rgb}{0.9961, 0.9062, 0.6328}
\definecolor{ieee-bright-yellow-20}{rgb}{0.9961, 0.9531, 0.8125}
\definecolor{ieee-bright-lgreen-100}{rgb}{0.4688, 0.7422, 0.125}
\definecolor{ieee-bright-lgreen-80}{rgb}{0.5742, 0.7852, 0.332}
\definecolor{ieee-bright-lgreen-60}{rgb}{0.6875, 0.8398, 0.5039}
\definecolor{ieee-bright-lgreen-40}{rgb}{0.793, 0.8906, 0.6641}
\definecolor{ieee-bright-lgreen-20}{rgb}{0.8945, 0.9414, 0.8281}
\definecolor{ieee-bright-dgreen-100}{rgb}{0.0, 0.5156, 0.2383}
\definecolor{ieee-bright-dgreen-80}{rgb}{0.1641, 0.6055, 0.3867}
\definecolor{ieee-bright-dgreen-60}{rgb}{0.3906, 0.6953, 0.5234}
\definecolor{ieee-bright-dgreen-40}{rgb}{0.6094, 0.8008, 0.6719}
\definecolor{ieee-bright-dgreen-20}{rgb}{0.8047, 0.8945, 0.8359}
\definecolor{ieee-bright-purple-100}{rgb}{0.5938, 0.1133, 0.5898}
\definecolor{ieee-bright-purple-80}{rgb}{0.6992, 0.3281, 0.668}
\definecolor{ieee-bright-purple-60}{rgb}{0.7812, 0.4961, 0.7461}
\definecolor{ieee-bright-purple-40}{rgb}{0.8555, 0.6602, 0.8281}
\definecolor{ieee-bright-purple-20}{rgb}{0.9219, 0.8281, 0.9023}
\definecolor{ieee-bright-lblue-100}{rgb}{0.0, 0.6094, 0.6484}
\definecolor{ieee-bright-lblue-80}{rgb}{0.0, 0.6797, 0.7188}
\definecolor{ieee-bright-lblue-60}{rgb}{0.2109, 0.75, 0.7812}
\definecolor{ieee-bright-lblue-40}{rgb}{0.5469, 0.8242, 0.8438}
\definecolor{ieee-bright-lblue-20}{rgb}{0.7695, 0.918, 0.9219}
\definecolor{ieee-bright-cyan-100}{rgb}{0.0, 0.707, 0.8828}
\definecolor{ieee-bright-cyan-80}{rgb}{0.0, 0.7227, 0.9453}
\definecolor{ieee-bright-cyan-60}{rgb}{0.2656, 0.7812, 0.957}
\definecolor{ieee-bright-cyan-40}{rgb}{0.5547, 0.8438, 0.9688}
\definecolor{ieee-bright-cyan-20}{rgb}{0.7773, 0.9141, 0.9805}
\definecolor{ieee-bright-white-100}{rgb}{0.9961, 0.9961, 0.9961}
\definecolor{ieee-bright-white-80}{rgb}{0.9961, 0.9961, 0.9961}
\definecolor{ieee-bright-white-60}{rgb}{0.9961, 0.9961, 0.9961}
\definecolor{ieee-bright-white-40}{rgb}{0.9961, 0.9961, 0.9961}
\definecolor{ieee-bright-white-20}{rgb}{0.9961, 0.9961, 0.9961}
\definecolor{ieee-dark-red-100}{rgb}{0.5234, 0.1211, 0.2539}
\definecolor{ieee-dark-red-80}{rgb}{0.6445, 0.2812, 0.3828}
\definecolor{ieee-dark-red-60}{rgb}{0.7422, 0.4727, 0.5234}
\definecolor{ieee-dark-red-40}{rgb}{0.832, 0.6445, 0.6758}
\definecolor{ieee-dark-red-20}{rgb}{0.918, 0.8203, 0.832}
\definecolor{ieee-dark-orange-100}{rgb}{0.9062, 0.4648, 0.1328}
\definecolor{ieee-dark-orange-80}{rgb}{0.9648, 0.5664, 0.3164}
\definecolor{ieee-dark-orange-60}{rgb}{0.9766, 0.6758, 0.4805}
\definecolor{ieee-dark-orange-40}{rgb}{0.9844, 0.7773, 0.6523}
\definecolor{ieee-dark-orange-20}{rgb}{0.9922, 0.8789, 0.8125}
\definecolor{ieee-dark-yellow-100}{rgb}{0.9961, 0.7773, 0.1719}
\definecolor{ieee-dark-yellow-80}{rgb}{0.9961, 0.8086, 0.375}
\definecolor{ieee-dark-yellow-60}{rgb}{0.9961, 0.875, 0.4492}
\definecolor{ieee-dark-yellow-40}{rgb}{0.9961, 0.8984, 0.6875}
\definecolor{ieee-dark-yellow-20}{rgb}{0.9961, 0.9453, 0.8438}
\definecolor{ieee-dark-lgreen-100}{rgb}{0.3945, 0.5508, 0.0938}
\definecolor{ieee-dark-lgreen-80}{rgb}{0.5078, 0.6289, 0.293}
\definecolor{ieee-dark-lgreen-60}{rgb}{0.6367, 0.7188, 0.4688}
\definecolor{ieee-dark-lgreen-40}{rgb}{0.7539, 0.8047, 0.6367}
\definecolor{ieee-dark-lgreen-20}{rgb}{0.875, 0.9023, 0.8125}
\definecolor{ieee-dark-dgreen-100}{rgb}{0.0, 0.3867, 0.2539}
\definecolor{ieee-dark-dgreen-80}{rgb}{0.1836, 0.5, 0.3906}
\definecolor{ieee-dark-dgreen-60}{rgb}{0.3984, 0.6172, 0.5273}
\definecolor{ieee-dark-dgreen-40}{rgb}{0.5938, 0.7422, 0.6758}
\definecolor{ieee-dark-dgreen-20}{rgb}{0.793, 0.8711, 0.8359}
\definecolor{ieee-dark-purple-100}{rgb}{0.4648, 0.1445, 0.5117}
\definecolor{ieee-dark-purple-80}{rgb}{0.5898, 0.3242, 0.6016}
\definecolor{ieee-dark-purple-60}{rgb}{0.6914, 0.4883, 0.6953}
\definecolor{ieee-dark-purple-40}{rgb}{0.7969, 0.6523, 0.793}
\definecolor{ieee-dark-purple-20}{rgb}{0.8945, 0.8203, 0.8945}
\definecolor{ieee-dark-cyan-100}{rgb}{0.0, 0.4492, 0.4648}
\definecolor{ieee-dark-cyan-80}{rgb}{0.0, 0.5469, 0.5664}
\definecolor{ieee-dark-cyan-60}{rgb}{0.3047, 0.6602, 0.668}
\definecolor{ieee-dark-cyan-40}{rgb}{0.5586, 0.7695, 0.7734}
\definecolor{ieee-dark-cyan-20}{rgb}{0.7734, 0.8789, 0.8789}
\definecolor{ieee-dark-dblue-100}{rgb}{0.0, 0.1562, 0.332}
\definecolor{ieee-dark-dblue-80}{rgb}{0.1797, 0.3008, 0.4609}
\definecolor{ieee-dark-dblue-60}{rgb}{0.3828, 0.4609, 0.5859}
\definecolor{ieee-dark-dblue-40}{rgb}{0.5781, 0.6289, 0.7188}
\definecolor{ieee-dark-dblue-20}{rgb}{0.7852, 0.8047, 0.8555}
\definecolor{ieee-dark-grey-100}{rgb}{0.457, 0.4688, 0.4805}
\definecolor{ieee-dark-grey-80}{rgb}{0.5625, 0.5625, 0.5742}
\definecolor{ieee-dark-grey-60}{rgb}{0.6641, 0.6641, 0.6758}
\definecolor{ieee-dark-grey-40}{rgb}{0.7734, 0.7695, 0.7773}
\definecolor{ieee-dark-grey-20}{rgb}{0.8789, 0.8828, 0.8828}
\definecolor{ieee-dark-black-100}{rgb}{0.0, 0.0, 0.0}
\definecolor{ieee-dark-black-80}{rgb}{0.3438, 0.3477, 0.3555}
\definecolor{ieee-dark-black-60}{rgb}{0.5, 0.5078, 0.5195}
\definecolor{ieee-dark-black-40}{rgb}{0.6523, 0.6602, 0.6719}
\definecolor{ieee-dark-black-20}{rgb}{0.8164, 0.8242, 0.8281}

\definecolor{light-gray}{gray}{0.75}

\newcommand{\etal}{\emph{et al.}}
\newcommand{\x}{$\times$}

\newcommand{\cmark}{\ding{51}}%
\newcommand{\xmark}{\ding{55}}%

\renewcommand{\subsubsection}[1]{\paragraph*{\textbf{#1}}}

\DeclareSIUnit{\x}{\!\ensuremath{\times}}
\DeclareSIUnit\bit{b}
\DeclareSIUnit\GE{GE}
\DeclareSIUnit\kGE{\kilo\GE}
\DeclareSIUnit\MGE{\mega\GE}
\ifx\showrevisionfromrejected\undefined
    \newcommand{\revrej}[1]{#1}
\else
    \newcommand{\revrej}[1]{{\textcolor{ieee-bright-red-100}{#1}}}
\fi

\ifx\showtvlsirebuttal\undefined
    \newcommand{\tvlsirev}[1]{#1}
\else
    \newcommand{\tvlsirev}[1]{{\textcolor{ieee-bright-lblue-100}{#1}}}
\fi

\ifx\showtvlsirebuttal\undefined
    \newcommand{\tvlsirevdel}[1]{\ignorespaces}
\else
    \newcommand{\tvlsirevdel}[1]{\textcolor{ieee-bright-red-100}{\st{#1}}}
\fi

\ifx\showtvlsirebuttal\undefined
    \newcommand{\tvlsirevrep}[2]{#2}
\else
    \newcommand{\tvlsirevrep}[2]{\tvlsirevdel{#1} \tvlsirev{#2}}
\fi

\ifx\showrebuttal\undefined
    \newcommand{\rev}[1]{#1}
\else
    \newcommand{\rev}[1]{{\textcolor{ieee-bright-lblue-100}{#1}}}
\fi

\ifx\showrebuttal\undefined
    \newcommand{\revdel}[1]{\ignorespaces}
\else
    \newcommand{\revdel}[1]{\textcolor{ieee-bright-red-100}{\st{#1}}}
\fi

\ifx\showrebuttal\undefined
    \newcommand{\revrep}[2]{#2}
\else
    \newcommand{\revrep}[2]{\revdel{#1} \rev{#2}}
\fi

\ifx\showrebuttal\undefined
    \newcommand{\revprg}[1]{\ignorespaces}
\else
    \newcommand{\revprg}[1]{\hspace{-0.5ex}\textcolor{ieee-bright-red-100}{\scalebox{.2}[1.5]{$\blacksquare$}}\hspace{-0.5ex}}
\fi

\ifx\showproof\undefined
    
\else
    
\fi

\ifx\showproof\undefined
    \newcommand{\proofdel}[1]{}
\else
    \newcommand{\proofdel}[1]{\textcolor{red}{#1}}
\fi

\ifx\showtvlsirebuttal\undefined
    \newcommand{\todo}[1]{}
\else
    \newcommand{\todo}[1]{{\textcolor{red}{#1}}}
\fi

\DeclareSIUnit\GE{GE}
\DeclareSIUnit\bit{b}
\DeclareSIUnit\kGE{\kilo\GE}
\DeclareSIUnit\MGE{\mega\GE}

\def\thetitle{ControlPULPlet: A Flexible Real-time Multi-core RISC-V Controller for~\revdel{Autonomous} 2.5D Systems-in-package}

\def\thetitleoneline{ControlPULPlet: A Flexible Real-time Multi-core RISC-V Controller for 2.5D Systems-in-package}

\bstctlcite{IEEEexample:BSTcontrol}

\begin{document}

\newcolumntype{R}{>{\raggedleft\arraybackslash}X}

\newacronym{dtm}{DTM}{dynamic thermal management}
\newacronym{dpm}{DPM}{dynamic power management}
\newacronym{dtpm}{DTPM}{dynamic thermal and power management}
\newacronym{stm}{STM}{static thermal management}
\newacronym{hw}{HW}{hardware}
\newacronym{sw}{SW}{software}
\newacronym{ca}{CA}{command/address}
\newacronym[plural={intellectual properties (IPs)}, firstplural={intellectual properties (IPs)}]{ip}{IP}{intellectual property}
\newacronym{ddr}{DDR}{double data rate}
\newacronym{sdr}{SDR}{single data rate}
\newacronym{lpddr}{LPDDR}{low-power double data rate}
\newacronym{rpc}{RPC}{reduced pin count}
\newacronym{dma}{DMA}{direct memory access}
\newacronym{axi}{AXI}{Advanced eXtensible Interface}
\newacronym{dram}{DRAM}{dynamic random access memory}
\newacronym[firstplural=static random access memories (SRAMs)]{sram}{SRAM}{static random access memory}
\newacronym{edram}{eDRAM}{embedded DRAM}
\newacronym[plural=SoCs, firstplural=systems on chip (SoCs)]{soc}{SoC}{system on chip}
\newacronym{mpsoc}{MPSoC}{multi-processor system on chip}
\newacronym{hesoc}{HeSoC}{heterogeneous system on chip}
\newacronym{fpga}{FPGA}{field-programmable gate array}
\newacronym{asic}{ASIC}{application-specific integrated circuit}
\newacronym{phy}{PHY}{physical layer}
\newacronym{ml}{ML}{machine learning}
\newacronym{iot}{IoT}{internet of things}
\newacronym{foss}{FOSS}{free and open source}
\newacronym{cmos}{CMOS}{complementary metal-oxide-semiconductor}
\newacronym{sut}{SUT}{system under test}
\newacronym{isut}{ISUT}{integrated system under test}
\newacronym{rtl}{RTL}{register transfer level}
\newacronym{hil}{HIL}{hardware in the loop}
\newacronym{pil}{PIL}{processor in the loop}
\newacronym{fil}{FIL}{FPGA in the loop}
\newacronym{mil}{MIL}{model in the loop}
\newacronym{sil}{SIL}{software in the loop}
\newacronym{hpc}{HPC}{high performance computing}
\newacronym{mcu}{MCU}{microcontroller unit}
\newacronym{fub}{FUB}{functional unit block}
\newacronym{dcu}{DCU}{domain control unit}
\newacronym{adas}{ADAS}{advanced driver-assistance system}
\newacronym{fame}{FAME}{FPGA Architecture Model Execution}
\newacronym{pl}{PL}{Programmable Logic}
\newacronym{ps}{PS}{Processing System}
\newacronym{apu}{APU}{Application Processing Unit}
\newacronym{ocm}{OCM}{on-chip memory}
\newacronym{pcs}{PCS}{power controller system}
\newacronym{pcf}{PCF}{power control firmware}
\newacronym{pmca}{PMCA}{programmable multi-core accelerator}
\newacronym{bram}{BRAM}{block RAM}
\newacronym{lut}{LUT}{look-up table}
\newacronym{ff}{FF}{flip-flop}
\newacronym{fsbl}{FSBL}{First Stage BootLoader}
\newacronym{pvt}{PVT}{Process, Voltage, Temperature}
\newacronym{hls}{HLS}{high-level synthesis}
\newacronym{mqtt}{MQTT}{Message Queuing Telemetry Transport}
\newacronym{cots}{COTS}{commercial off-the-shelf}
\newacronym{cpu}{CPU}{central processing unit}
\newacronym{gpu}{GPU}{graphic processing unit}
\newacronym{ibmocc}{IBM OCC}{IBM on-chip controller}
\newacronym{clic}{CLIC}{core-local interrupt controller}
\newacronym{clint}{CLINT}{Core-Local Interruptor}
\newacronym{scmi}{SCMI}{System Control and Management Interface}
\newacronym{os}{OS}{Operating System}
\newacronym{ospm}{OSPM}{operating system-directed configuration and power management}
\newacronym{mimo}{MIMO}{Multiple-Input Multiple-Output}
\newacronym{siso}{SISO}{Single-Input Single-Output}
\newacronym{bmc}{BMC}{Baseboard Management Controller}
\newacronym{qos}{QoS}{quality of service}
\newacronym{tdp}{TPD}{thermal design power}
\newacronym{dvfs}{DVFS}{dynamic voltage and frequency scaling}
\newacronym{dfs}{DFS}{dynamic frequency scaling}
\newacronym{dvs}{DVS}{dynamic voltage scaling}
\newacronym{rtu}{RTU}{Real Time Unit}
\newacronym{pe}{PE}{processing element}
\newacronym{noc}{NoC}{network on chip}
\newacronym{noi}{NoI}{network on interposer}
\newacronym{pid}{PID}{proportional integral derivative}
\newacronym{sota}{SoA}{state-of-the-art}
\newacronym{fpu}{FPU}{floating point unit}
\newacronym{pcu}{PCU}{Power Control Unit}
\newacronym{scp}{SCP}{System Control Processor}
\newacronym{mcp}{MCP}{Manageability Control Processor}
\newacronym{occ}{OCC}{On-Chip Controller}
\newacronym{smu}{SMU}{System Management Unit}
\newacronym{ap}{AP}{application-class processor}
\newacronym{vrm}{VRM}{voltage regulator module}
\newacronym{pfct}{PFCT}{periodic frequency control task}
\newacronym{pvct}{PVCT}{periodic voltage control task}
\newacronym{ipc}{IPC}{instructions per cycle}
\newacronym{simd}{SIMD}{single instruction, multiple data}
\newacronym{mctp}{MCTP}{Management Component Transport Protocol}
\newacronym{pldm}{PLDM}{Platform Level Data Model}
\newacronym{rtos}{RTOS}{real-time OS}
\newacronym{hlc}{HLC}{high-level controller}
\newacronym{llc}{LLC}{low-level controller}
\newacronym{acpi}{ACPI}{Advanced Configuration and Power Interface}
\newacronym{pdn}{PDN}{Power Delivery Network}
\newacronym{ewma}{EWMA}{Exponential Weight Moving Average}
\newacronym{ppa}{PPA}{power, performance and area}
\newacronym{pcb}{PCB}{Printed Circuit Board}
%
\newacronym{dsa}{DSA}{domain-specific accelerator}
\newacronym{ha}{HA}{Hardware Accelerator}

\newacronym[longplural={high-bandwidth memories}]{hbm}{HBM}{high-bandwidth memory}
\newacronym{rapl}{RAPL}{Running Average Power Limit}
\newacronym{tsv}{TSV}{through silicon via}
\newacronym{fet}{FET}{field effect transistor}
\newacronym{fll}{FLL}{frequency locked loop}
\newacronym{pll}{PLL}{phase locked loop}

\newacronym{oca}{EBA}{Enhanced Baseline Algorithm}
\newacronym{fca}{FCA}{Fuzzy-ispired Iterative Control Algorithm}
\newacronym{vba}{VBA}{Voting Box Algorithm}

\newacronym{pm}{PM}{power management}
\newacronym{pmi}{PMI}{power management interface}
\newacronym{fw}{FW}{firmware}
\newacronym{opal}{OPAL}{OpenPower abstraction layer}
\newacronym{pmbus}{PMBUS}{Power Management Bus}
\newacronym{avsbus}{AVSBUS}{Adaptive Voltage Scaling}
\newacronym{psci}{PSCI}{Power State Coordination Interface}
\newacronym{uefi}{UEFI}{Unified Extensible Firmware Interface}
\newacronym{asl}{ASL}{ACPI source language}
\newacronym{aml}{AML}{ACPI machine language}
\newacronym{msr}{MSR}{model-specific register}
\newacronym{mpc}{MPC}{model predictive control}
\newacronym{qp}{QP}{quadratic programming}
\newacronym{fp32}{FP32}{\texttt{float32}}
\newacronym{fp16}{FP16}{float16}
\newacronym{fp64}{FP64}{float64}
\newacronym{gp}{GP}{general-purpose}
\newacronym{ds}{DS}{domain-specific}
\newacronym{spm}{SPM}{scratchpad memory}
\newacronym{flops}{FLOPS}{floating-point operations per second}
\newacronym{rac}{RAC}{runtime active control}
\newacronym{rl}{RL}{reinforcement learning}
\newacronym{etm}{ETM}{energy and thermal management}
\newacronym{uav}{UAV}{unmanned aerial vehicles}
\newacronym{pulp}{PULP}{parallel ultra-low power}
\newacronym{osqp}{OSQP}{operator-splitting quadratic programming}
\newacronym{admm}{ADMM}{alternating direction method of multipliers}
\newacronym{isa}{ISA}{instruction set architecture}
\newacronym{vlsi}{VLSI}{very large scale integration}
\newacronym{deepc}{DeePC}{data-enabled predictive control}
\newacronym{sssr}{SSSR}{sparse stream semantic register}
\newacronym{dsp}{DSP}{digital signal processing}
\newacronym{kkt}{KKT}{Karush-Kuhn-Tucker}
\newacronym{amd}{AMD}{approximate minimum degree}
\newacronym{lqr}{LQR}{linear quadratic regulator}
\newacronym{lti}{LTI}{linear time-invariant}
\newacronym{fg}{FG}{fast gradient}
\newacronym{pcg}{PCG}{preconditioned conjugate gradient}
\newacronym{lns}{LNS}{logarithmic number system}
\newacronym{dmp}{DMP}{discrete model pruning}
\newacronym{ge}{GE}{gate equivalent}
\newacronym{axi4}{AXI4}{Advanced eXtensible Interface 4}
\newacronym{amba}{AMBA}{Advanced Microcontroller Bus Architecture}
\newacronym{sl}{SL}{serial link}
\newacronym{d2d}{D2D}{die-to-die}
\newacronym{mcm}{MCM}{multi-chip module}
\newacronym{cppc}{CPPC}{collaborative processor performance control}
\newacronym{obi}{OBI}{open bus interface}
\newacronym{axis}{AXIS}{AXI stream}
\newacronym{sip}{SiP}{system in package}
\newacronym{sips}{SiPs}{systems in package} 
\newacronym{pnr}{PnR}{place-and-route}
\newacronym{cdc}{CDC}{clock domain crossing}
\newacronym{cg}{CG}{clock gate}
\newacronym{isr}{ISR}{interrupt service routine}
\newacronym{qfn}{QFN}{quad-flat no-leads}
\newacronym{scf}{SCF}{scalable control fabric}

\linenumbersep 3pt\relax


\title{\thetitle}

\ifx\blind\undefined
    \author{
        Alessandro~Ottaviano~\orcidlink{0009-0000-9924-3536},~\IEEEmembership{Student Member, IEEE}, %
        Robert Balas~\orcidlink{0000-0002-7231-9315},~\IEEEmembership{Student Member, IEEE}, %
        Tim Fischer~\orcidlink{0009-0007-9700-1286},~\IEEEmembership{Student Member, IEEE}, %
        Thomas Benz~\orcidlink{0000-0002-0326-9676},~\IEEEmembership{Student Member, IEEE}, %
        Andrea Bartolini~\orcidlink{0000-0002-1148-2450},~\IEEEmembership{Member, IEEE},\\ %
        Luca Benini~\orcidlink{0000-0001-8068-3806},~\IEEEmembership{Fellow, IEEE} %
        \thanks{A.~Ottaviano, and R.~Balas  contributed equally to this work.}
        \IEEEcompsocitemizethanks{\IEEEcompsocthanksitem A.~Ottaviano, R.~Balas,  T.~Fischer, T.~Benz, and L.~Benini are with the Integrated Systems Laboratory (IIS), ETH Zurich, Switzerland.\protect\\
        E-mail: \{aottaviano,balasr,tbenz,fischeti,lbenini\}@iis.ee.ethz.ch
        \IEEEcompsocthanksitem L.~Benini is also with the Department of Electrical, Electronic and Information Engineering (DEI), University of Bologna, Bologna, Italy.\protect
        \IEEEcompsocthanksitem A.~Bartolini is with the Department of Electrical, Electronic and Information Engineering (DEI), University of Bologna, Bologna, Italy.\protect\\
        E-mail: a.bartolini@unibo.it
        }
    }
\else
    \author{%
            \emph{Authors omitted for blind review}
            }
\fi

\markboth{}%
{Ottaviano \MakeLowercase{\emph{et al.}}: \thetitleoneline}

\maketitle

\begin{abstract}
The growing complexity of real-time control algorithms with increasing performance demands, along with the shift to 2.5D technology, drive the need for scalable controllers to manage chiplets' coupled operation in 2.5D systems-in-package.
These controllers must offer real-time computing capabilities, as well as \revrep{chiplet-compatible}{System-in-package (SiP) compatible} IO interfaces for communicating with the controlled dies.
Due to real-time constraints, a key challenge is minimizing the performance penalty \revrep{of the chiplet paradigm compared to the inherently superior efficiency of}{of die-to-die communication with respect to} native on-chip control interfaces.
We address this challenge with \emph{ControlPULPlet}, \revrep{a chiplet-compatible}{an open-source}, real-time multi-core RISC-V controller \revrep{, which we release open-source}{designed specifically for SiP integration}.
\revrep{It}{ControlPULPlet} features a 32-bit CV32RT core for fast interrupt handling and a specialized direct memory access engine to automate periodic sensor readout.
A tightly-coupled programmable multi-core \revrep{accelerator}{cluster for acceleration of advanced control algorithms} is integrated through a dedicated AXI4 port.
A flexible AXI4-compatible die-to-die (D2D) link enables \revrep{inter-chiplet}{efficient} communication in 2.5D \revrep{systems}{SiPs.}~\revdel{ and high-bandwidth transfers in conventional 2D monolithic dies.}
We implemented and fabricated ControlPULPlet as a silicon demonstrator called \emph{Kairos} in TSMC's 65nm CMOS.
Kairos runs \rev{model} predictive control algorithms at up to \SI{290}{\mega\hertz} in a \SI{30}{\milli\watt} power envelope.
The D2D link attains a peak duplex transfer rate of \SI{51}{\giga\bit\per\second} at \SI{200}{\mega\hertz}, at the minimal costs of just \SI{7.6}{\kGE} in PHY area per channel, adding just \SI{2.9}{\percent} to the total system area.
\end{abstract}


%
%
%
%
%
%

\section{Introduction}

\begin{figure}[t]
    \centering
    \includegraphics[width=\columnwidth]{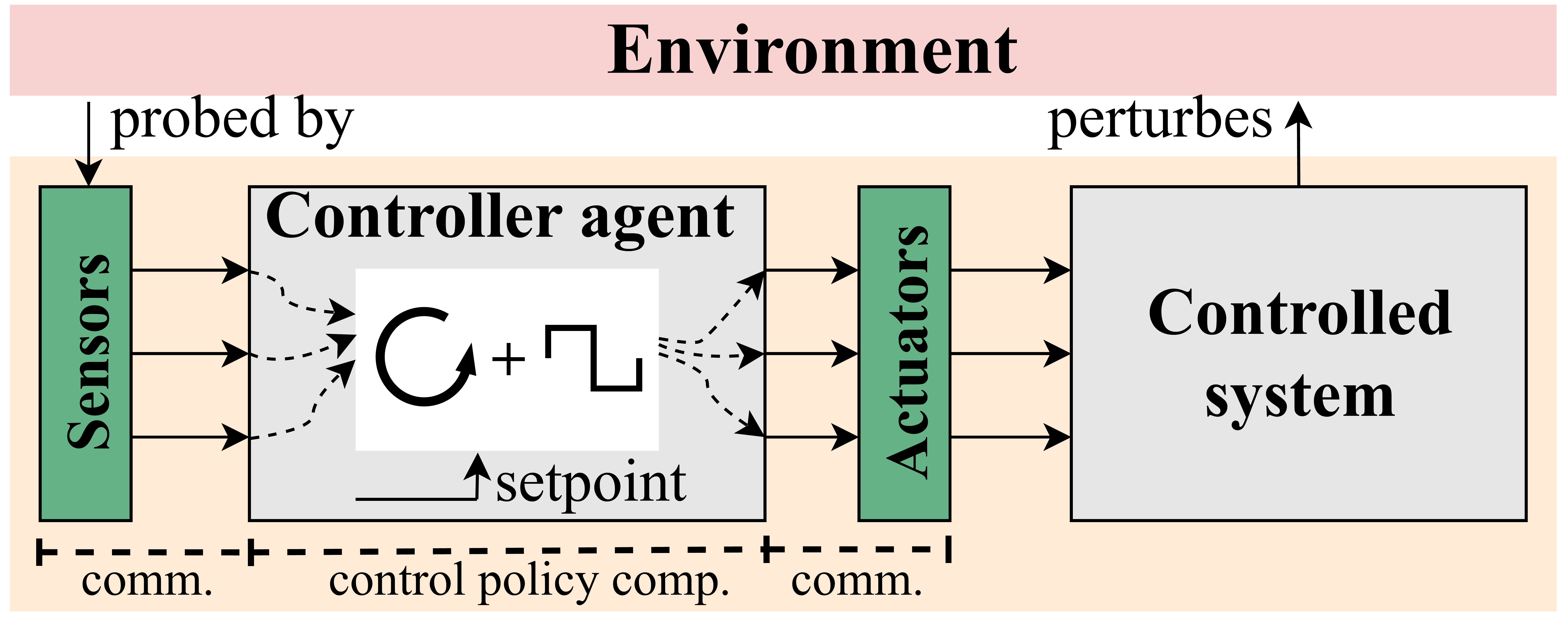}
    \caption{%
        Main elements involved in the control loop pipeline of a controller agent.
    }
    \label{fig:auto-sys}
\end{figure}

An increasing number of integrated systems \revdel{designed for unmanned, autonomous operations} rely on closed-loop control to meet their mission objectives \rev{in terms of performance, power consumption, and thermal stability}.
The control loop consists of three main components, depicted in~\cref{fig:auto-sys}. First, sensory circuits on the left side probe the current state of the system. 
Next, a control policy minimizes tracking errors while maintaining stability by closely following the setpoint of an objective function.
Finally, the resulting control actions are applied to physical actuators on the right side, leading to \revdel{proportional} changes in the controlled system's behavior and, therefore, perturbations within its environment.
Since the environment state evolves independently of the control pipeline, it is periodically probed by the sensory circuitry. Smaller sampling periods lead to finer granularity of control actions.
Such control systems are used across various application domains, including automotive systems and robotics~\cite{AUTOMOTIVE_PREDICTABLE}, power conversion~\cite{GANSI_1}, and CPU power management~\cite{Ottaviano2024,CONTROLLER_UMBRELLA_TVLSI_1}.
For instance, \revrep{in robotics, the reference objective function may represent a trajectory}{in a smart power converter, the objective function may be conversion efficiency}, while in CPU power management, it could correspond to a power budget based on runtime workload and temperature.

The system component executing the \revdel{se} control schemes is the \emph{controller agent}, or controller.
To maximize flexibility, controllers are implemented as embedded, \gls{mimo} digital programmable units (see \cref{sec:rel}), with power envelopes from tens to hundreds of \si{\milli\watt} and on-chip memory footprint under a few \si{\mebi\byte}~\cite{ARM_PCSA,Ottaviano2024}.

As shown in \cref{fig:auto-sys}, a controller performs two main tasks: data \emph{communication} during sensor readout and actuator \revrep{control}{configuration}, and \emph{processing} for control policy computation. \revrep{In autonomous systems, t}{T}hese tasks are constrained by three key factors: predictability of execution, real-time performance, and reliability. This work focuses on the first two properties\revdel{, as reliability forms a distinct line of research}.
%
Predictable and real-time execution means the communication and processing pipelines must complete deterministically before the next control loop period begins, known as the deadline.
\revrep{There are three ways to achieve this}{To achieve this goal, we need to meet three design objectives}: 
\textbf{(1)} designing controllers with predictable hardware features, such as cache-less memory hierarchies using \glspl{spm} or simple, in-order, non-speculative processor cores;
\textbf{(2)} ensuring physical isolation between the controller and its environment to reduce interference on shared resources, such as the interconnect (bus) hierarchy. 
Isolation includes runtime freedom from interference, fabrication-level independence, and standalone certification for safety integrity~\cite{MCS_REVIEW_1};
\textbf{(3)} minimizing communication and processing latency.
Communication latency can be reduced through high-bandwidth interfaces between the controller and the controlled system, and through fast interrupt response times ($<$ 10 clock cycles). 
Processing latency can be improved by scaling the controller's computing resources (e.g., \glspl{dsa} or multi-core architectures) tailored to the complexity of the control policy.

\begin{figure}[t]
    \centering
    \includegraphics[width=\columnwidth]{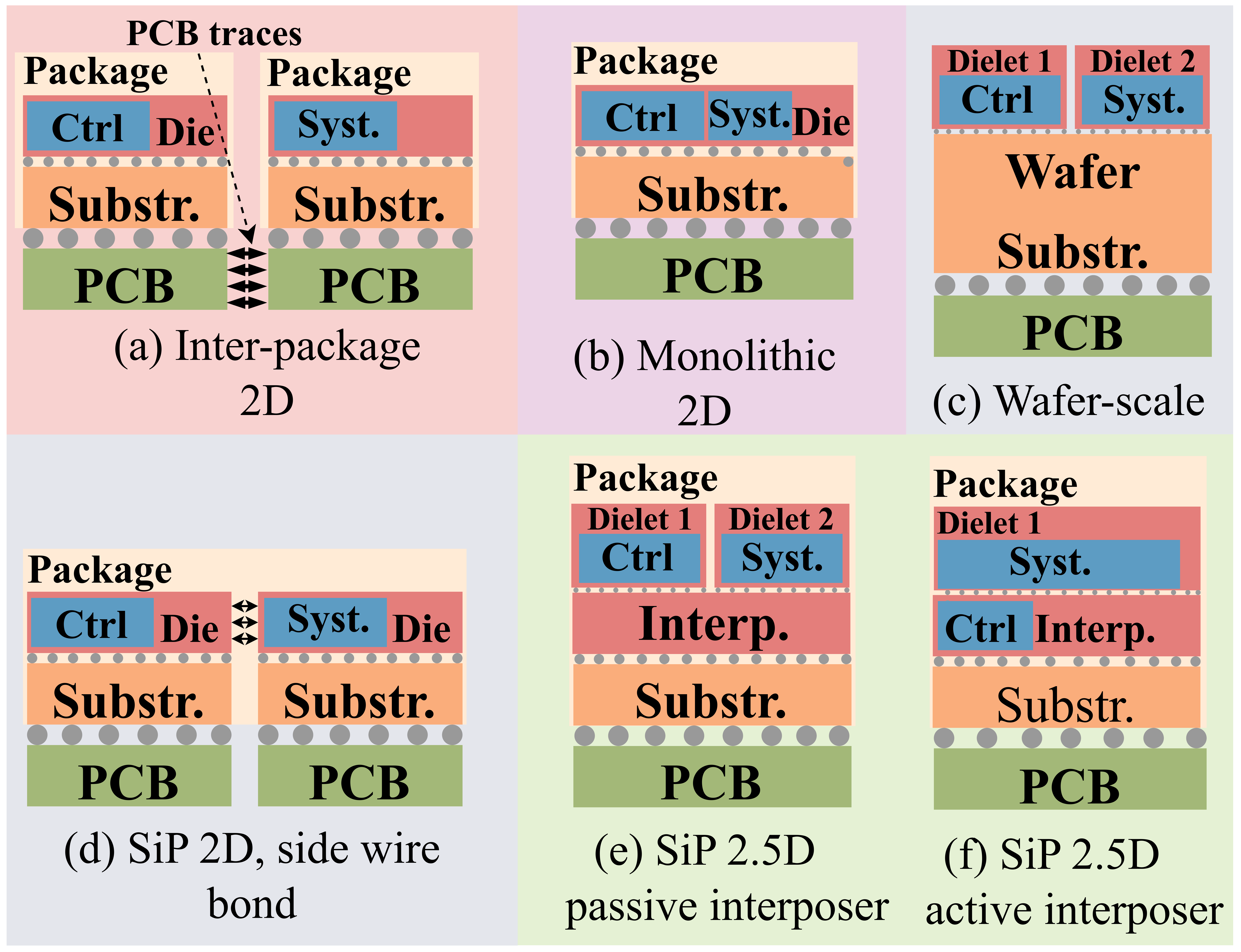}
    \caption{%
        Controller topologies, categorized by packaging technology.
    }
    \label{fig:ctrl-types}
\end{figure}

Another impacting factor on the isolation and communication latency is the integration strategy of the controller and the controlled system, which we depict in~\cref{fig:ctrl-types}.
Inter-package integration (\cref{fig:ctrl-types}a) offers excellent isolation but higher communication latency. 
Conversely, monolithic integration on the same silicon die (\cref{fig:ctrl-types}b) minimizes latency but \revrep{compromises}{hinders} isolation, as components share resources during operation and are logically indivisible as part of the same fabrication process.

\Gls{sips} offer a tradeoff between isolation and latency by enclosing multiple chips in a single carrier.
In this setup, \gls{d2d} connections between the controller and the system can be implemented using wire bonding (\cref{fig:ctrl-types}d), as seen in commercial Si-GaN power switches~\cite{GAN_PACKAGE}.
The latest development in \gls{sip} technology is the "disintegration" of a monolithic silicon die in many \emph{chiplets}, or dielets.
This post-Moore approach reduces design and production costs, improves production yield thanks to smaller chip sizes, and facilitates the integration of heterogeneous components fabricated independently~\cite{ISSCC_TRENDS_2023}. 
%
%
Passive or active interposers hosting \gls{d2d} connectivity fabrics~\cite{DasSharmaNature2024} offer up to $44 \times$ the bandwidth of off-chip links~\cite{HEXAMESH}, though still lower than native on-die communication.
\Cref{fig:ctrl-types}e-f show an example of controller integration with passive and active interposers, respectively.
\tvlsirev{Beyond interposer technology, bleeding-edge technologies like wafer-scale integration enable dense, low-latency chiplet communication via redistribution layers or silicon bridges on wafer, bypassing traditional packaging, as shown in~\cref{fig:ctrl-types}c.}

Early adopters of chiplet technology in commercial products, such as AMD, have leveraged it in \gls{hpc} systems \tvlsirev{and data-intensive \gls{ml} workloads} to address the longer development cycles and increased expenses of transitioning to advanced manufacturing nodes in conventional monolithic integration~\cite{AMD_ISCA_PIONEERING,AMD_DATE21,CHIPLET_1}.
%
Recently, chiplets have been explored and prototyped in other domains, e.g., Si-GaN switches, to achieve tighter integration between controllers and controlled systems~\cite{GANSI_1}. 
\revdel{These systems, though less complex than high-performance designs, leverage legacy technology nodes and only require specialized functions.}

\revrep{Conversely, programmable}{Most} \tvlsirev{embedded} controllers with real-time and predictable execution requirements continue to rely on monolithic integration. 
\revdel{This is because their application domains, traditionally outside the high-performance spectrum, have not yet encountered the scalability, modularity, and cost barriers that have driven the adoption of the chiplet paradigm in high-performance computing systems.} 
However, as advanced autonomous systems --- such as humanoid robots and self-driving cars --- demand \revrep{increasingly higher}{growing} computational performance \rev{and integration,} while maintaining real-time constraints, they are beginning to face similar challenges \rev{to high-performance systems}, making \revrep{chiplet-based architectures a}{SiP approaches for control integration a} compelling alternative.

\revdel{Additionally, another advantage of chiplet technology is its ability to provide enhanced isolation by design. 
Dedicated manufacturing of standalone silicon dies ensures functional separation from other subsystems during operation and simplifies certification processes for time- and safety-critical applications. 
Some recent prototypes in the automotive domain are beginning to explore these possibilities
However, careful consideration must be given to the communication overhead introduced by the chiplet paradigm, as it may impact real-time performance and predictability constraints.}


In light of these opportunities and challenges, we present ControlPULPlet, a \revrep{programmable multi-core controller for seamless integration into conventional on-chip dies or as a chiplet for \glsentrytext{sips}}{platform that can be configured for monolithic or \glsentrytext{sip} integration}.
The platform builds on the RISC-V ControlPULP controller architecture, which follows the programmable multi-core accelerator (PMCA) paradigm~\cite{Ottaviano2024}. This paper introduces several \tvlsirev{hardware} enhancements compared to the ControlPULP architecture \tvlsirev{towards a flexible, real-time embedded controller with 2.5D interface compatibility}: (i) \tvlsirevrep{it proposes a D2D control interface that can be seamlessly selected as an alternative to the native AXI4 interface as the main connection point towards the controlled system}{it introduces a D2D control interface that can be used as a drop-in alternative to the native AXI4 interface for connecting to the controlled system}, facilitating integration in chiplet designs; (ii) it extends the general-purpose core functionality with fast interrupt handling through a lightweight ISA extension, essential in real-time systems, and a specialized, autonomous direct memory access (DMA) engine for \tvlsirevrep{efficient}{periodic} \tvlsirevdel{sensor} data acquisition \tvlsirevdel{in periodic control loops} through the D2D link \tvlsirev{without any periodic intervention of the processor}; (iii) finally, it evaluates these additional features on a silicon demonstrator chip, called \emph{Kairos}.
The synthesizable hardware description is \ifx\blind\undefined
     open-source \rev{under a liberal license}\footnote{\revrej{\texttt{\url{https://github.com/pulp-platform/control-pulp/}}}}.
\else
     open-source \rev{under a liberal license}.\footnote{\texttt{URL omitted for blind review}}
\fi

This paper makes the following contributions:
\begin{itemize}
    \item An open-source, embedded RISC-V controller with comprehensive real-time and computing capabilities compatible with the monolithic and chiplet design paradigms. \revrep{As previously mentioned, t}{T}he design enhances the open-source ControlPULP~\cite{Ottaviano2024}, tuned for on-chip control applications~(\cref{sec:archi}).
    
    \item A scalable, AXI4-compatible source-synchronous digital \gls{d2d} link. 
    With eight channels and a flow control buffer depth of 128, the link achieves an average peak bus utilization of \SI{85}{\percent}, compared to \SI{95}{\percent} for its on-chip equivalent~(\cref{subsec:eval:bw_lat}). 
    This setup incurs a minimal PHY area overhead of \SI{2.9}{\percent} and results in a negligible performance impact \tvlsirevrep{in the free time window of a MIMO}{to the latency of} \tvlsirevdel{periodic} \tvlsirev{an exemplary} control loop application (\cref{subsec:eval:case_study}).
    Increasing the buffer depth enhances utilization and matches on-chip control performance, albeit at additional area cost. 

    \item Integration of hardware enhancements for real-time execution: fast interrupt handling and context switching~\cite{balas2023cv32rt}, and \gls{dma} with periodic mid-end~\cite{benz2023highperformance}~(\cref{subsec:archi:rt}).
    
    \item A standalone, single-core demonstrator chip called \emph{Kairos}, fabricated in TSMC's \SI{65}{\nano\meter} node. At \SI{1.2}{\volt}, Kairos achieves a peak clock frequency of \SI{290}{\mega\hertz} with a power envelope not exceeding \SI{30}{\milli\watt} during data-intensive control workloads~(\cref{sec:eval}). On the chip, the \gls{d2d} link attains a duplex peak transfer rate of \SI{51.2}{\giga\bit\per\second} at the nominal \SI{200}{\mega\hertz} (\cref{sec:eval}).
\end{itemize}

\section{Architecture}\label{sec:archi}

\begin{figure}[t]
    \centering
    \includegraphics[width=\columnwidth]{fig/cpulplet.pdf}
    \caption{%
        \revrej{ControlPULPlet architecture, building upon the work in~\cite{Ottaviano2024}. Extensions introduced in this work are highlighted in color, with pre-existing architectural blocks shown in grayscale for distinction. \textcolor{ieee-bright-lgreen-60}{$\blacksquare$} represents a bus adapter that convert between \gls{obi} and AXI4 protocols, and between \SI{32}{\bit} and \SI{64}{\bit} bus widths.}
    }
    \label{fig:cpulplet-archi}
\end{figure}

In this section, we first describe the ControlPULPlet platform, highlighting the main enhancements compared to the \revrep{original design}{ControlPULP architecture} (\cref{subsec:archi:overview}).
Then, we dive into the integration of the real-time features (\cref{subsec:archi:rt}) and the architecture of the \gls{d2d} link (\cref{subsec:archi:d2d}).

\subsection{ControlPULPlet Platform}\label{subsec:archi:overview}

The high-level block diagram of the ControlPULPlet controller is illustrated in~\cref{fig:cpulplet-archi}. 
Pre-existing architectural blocks are represented in grayscale, while architectural additions introduced in this work are highlighted with a distinct color scheme for clarity and differentiation.

\subsubsection{Manager and \gls{pmca} Domains}
The design is centered around a 32-bit, RISC-V \emph{manager domain} integrating a CV32RT in-order core~\cite{balas2023cv32rt}.
%
To accelerate compute-intensive control workloads, such as \gls{mpc}, an 8-core, \gls{dma}-capable \gls{pmca} with per-core \SI{32}{\bit} \glspl{fpu} connects to the manager domain; \revrej{the \gls{pmca} has been extensively validated in~\cite{Ottaviano2024}.}
Communication between these domains occurs via two \SI{64}{\bit} AXI4 \gls{dsa} ports, shown in~\cref{fig:cpulplet-archi}.
These ports can be disabled at design time for simpler applications where a single core suffices.

\subsubsection{Interconnect System}
The interconnect fabric consists of \SI{32}{\bit} data and address-wide, 2-clock-cycle access latency, \gls{obi}-compliant network elements for low and predictable access latency to the memory system.
To bridge with the on-chip and inter-chiplet IO interface, detailed below, the \SI{32}{\bit} \gls{obi}-compliant data interface is converted to a \SI{64}{\bit} AXI4-compliant interface, and arbitrated by an AXI4 crossbar with 3-clock-cycles latency.
Several bus adapters convert between the two protocols\tvlsirevdel{ and their bus widths}, as shown in~\cref{fig:cpulplet-archi} by the small green rectangles.

\subsubsection{On-chip and Inter-chiplet IO Interface}

The subordinate interface of the \SI{64}{\bit} AXI4 bus --- indicated as \emph{sAXI4} in~\cref{fig:cpulplet-archi} --- is responsible for booting the controller. This can be performed by the controlled system, or a separate, secure hardware root-of-trust.

The manager interface --- indicated as \emph{mAXI4} in~\cref{fig:cpulplet-archi} --- realizes the sensing/actuation communication path shown in~\cref{fig:auto-sys}. It is responsible for sensor readout and dispatch of the quantities produced by the control policy.
As illustrated in~\cref{fig:cpulplet-archi}, a bypass network utilizing AXI4-compatible network switches enables selection between the on-chip AXI4 interface and the \gls{d2d} protocol for I/O connectivity.
This selection occurs at the AXI4 bus level, which serves as the frontend network layer of the \gls{d2d} link. The detailed microarchitecture of the latter is provided in \cref{subsec:archi:d2d}.
The two interfaces are mutually exclusive, catering to distinct use cases: native AXI4 facilitates on-chip integration with controlled systems, while the \gls{d2d} protocol supports chiplet-based integration.
Consequently, the bypass configuration is static and must be determined at design time prior to physical implementation.

A \SI{64}{\bit}, AXI-compliant system \gls{dma} engine connects directly with the outgoing \gls{d2d} interface to decouple off-chiplet transfers from other tasks running on the manager and \gls{pmca} domains.
\tvlsirev{This separation also ensures that high-contention dataplane transfers do not interfere with time-critical control-plane operations~\cite{PWRMGMT_INTEL,AMD_SMU}, which are the primary focus of this work.}
\tvlsirev{The AXI4-compliant interface enables high-performance transfers through support for burst transactions and multiple outstanding requests.}
The \gls{dma} engine integrates a hardware extension for autonomous, core-independent data movement, which we detail in~\cref{subsec:archi:rt}. 
To allow also the manager core to interact with the controlled system, an additional AXI4 manager port is arbitrated with the system DMA port (green rectangle in~\cref{fig:cpulplet-archi}).

\subsubsection{Memory System}
An on-chip, \SI{512}{\kibi\byte} shared L2 \gls{spm} in the manager domain serves as the main exchange point with the \glspl{pmca}, which owns a private, bank-interleaved \SI{128}{\kibi\byte} L1 \gls{spm}. 
In both domains, the \gls{obi} interconnect ensures constant access times to the memory endpoints.
To minimize access conflicts between large burst transfers from the system or \gls{dsa} \gls{dma} engines and narrow transfers from the manager core, the L2 \gls{spm} banks use an interleaved addressing scheme. The L1 \gls{spm} employs a similar scheme to reduce interference among the 8 \gls{pmca} cores.

\subsubsection{Doorbell-based Mailboxes Messaging}
ControlPULPlet features a configurable doorbell-based mailbox unit with 64 default mailboxes, enabling real-time communication for dispatching requests, setpoint updates, or control policy constraints.
The mailboxes are accessible through the AXI4 subordinate external port from the controlled system.
\tvlsirevdel{Popular protocols such as Arm's SCMI, the in-progress RISC-V platform management interface (RPMI), or DDS-XRCE can be used as software abstractions to manage the message exchange.}
A RISC-V \gls{clic} handles incoming interrupt lines from peripherals and the mailbox unit; in the latter case, one interrupt line per mailbox is used.

\subsubsection{Off-chip IO Interface}
To handle off-chip communication with the controlled system, as in the case of discrete physical actuators like motors, brakes, robotic arms, and voltage regulator modules, ControlPULPlet supports a broad set of \gls{sota} peripherals.
Each IO peripheral is managed by a private, IO \gls{dma}, featuring a configurable number of peripherals, defaulting to 32 GPIOs, 1 UART, 12 I2C, and 8 SPI.
The \gls{dma} core architecture mirrors that used for on-chip or \gls{d2d} communication, since the unit supports multiple interface protocols and thus enables the same hardware to handle diverse scenarios~\cite{benz2023highperformance}. 
The autonomous data transfer extension is also supported for off-chip peripherals.
Finally, two \SI{32}{\bit} system timers and a PWM timer enable diverse periodic control scenarios, such as motor control.

\subsubsection{\revrej{Programming Model}}

ControlPULPlet employs a lightweight \gls{rtos}-based programming model, supporting \tvlsirevrep{various}{multiple} \glspl{rtos}\tvlsirevdel{ such as FreeRTOS, ThreadX, and Zephyr}. 
An \gls{rtos} \tvlsirevrep{enables the creation of periodic tasks managed at the firmware level using platform timers, essential for handling periodic control applications}{manages periodic tasks in firmware via platform timers, essential for control applications}. 
This approach underscores two key aspects. First, it emphasizes \tvlsirevrep{the importance of fast interrupt handling extensions to optimize context switching, a very common mechanism when a single-core processor handles different tasks simultaneously, as we discuss in Section II-B}{fast interrupt handling to optimize context switching on single-core processors (\cref{subsec:archi:rt})}.
Second, it highlights \tvlsirevrep{the advantages of a fully programmable controller that balances flexibility and specialization by leveraging the open RISC-V ISA}{the benefits of a programmable RISC-V-based controller that balances flexibility and specialization}.
Such a controller \tvlsirevrep{can integrate with a variety of standardized RTOSs --- tailored to meet diverse safety and time-critical requirements --- and can thereby be employed for various control applications}{supports standardized RTOSs, making it suitable for diverse safety- and time-critical control applications}.

\subsection{\revrej{Real-time features}}\label{subsec:archi:rt}

\revrej{\revrep{The following paragraphs outline}{We detail} the real-time extensions integrated into the ControlPULPlet platform, focusing on the interrupt subsystem and autonomous sensor readout. 
The final paragraph details their system-level integration.}

\subsubsection{\revrej{CV32RT}} 
The CV32RT core achieves exceptional performance for real-time tasks through several architectural innovations tailored for minimizing interrupt latency and context switching time.
\tvlsirev{It integrates a fast interrupt controller based on the Core Local Interrupt Controller (CLIC) RISC-V standard.}
The CLIC introduces support for vectored interrupts, level-based prioritization, and selective hardware vectoring (SHV), which accelerates both nested and non-nested interrupt scenarios. \tvlsirevdel{These features surpass the SoA RISC-V interrupt controller, the CLINT.}
Additionally, the \tvlsirevdel{integration of the} \texttt{fastirq} \tvlsirev{lightweight ISA} extension \tvlsirev{of the core} reduces the interrupt latency and context switching to as low as 6 and 100 fixed clock cycles, respectively, \tvlsirevrep{by utilizing bank switching and a background context-saving mechanism.}{on-par to SoA at low area cost. Therefore, the RTOS overhead is bound to 107 clock cycles for context switch and timer tick setup (1 cycle).}
\tvlsirev{\texttt{fastirq}'s architecture comprises a dual-banked register file with an integrated FSM that triggers a background, fully hardware-managed context-saving mechanism upon interrupt arrival. A bank switch allows immediate execution of the handler while the prior context is asynchronously stored to memory via a dedicated datapath. The design includes stack pointer adjustment logic and memory access coordination to avoid conflicts, enabling consistent execution even during partial saves.} 
\tvlsirev{When equal-priority, back-to-back interrupts are fired,} \textit{tail-chaining} minimizes redundant context save/restore sequences, \tvlsirevrep{improving the response time for back-to-back interrupts.}{preserving only the first save and the final restore.}

\subsubsection{\revrej{DMA With Real-time Mid-end}} 

\revrej{For real-time data acquisition tasks, such as reading sensor arrays with complex and irregular address maps, the system integrates a modular \gls{dma}~\cite{benz2023highperformance}.
A dedicated real-time (RT) mid-end \tvlsirevrep{enables repeated N-dimensional transfers with highly configurable parameters, such as periodicity and shape, controlled via software. The decoupled control and data planes enable parallelization of computation and data transfer, reducing overhead on general-purpose cores.}{manages N-dimensional data transfers with support for highly configurable parameters such as stride, shape, and period, enabling precise control of streaming patterns via software.
The mid-end architecture includes a microcoded transfer controller and a programmable sequencer, which orchestrate address generation and handshake logic independently from the data path. This decoupling of control and data planes allows data movement to proceed in parallel with computation, minimizing contention on shared interconnects and offloading the general-purpose cores. A loop flattening unit and nested loop counter enable complex multidimensional access patterns, while internal FIFOs and backpressure-aware arbitration logic ensure timing-accurate execution in real-time workloads.}
\tvlsirevdel{This architecture allows for autonomous, hardware-accelerated memory operations, ensuring deterministic periodicity and low latency while supporting complex, high-dimensional data access patterns.} 
\tvlsirevdel{This capability}{The autonomous DMA}  is crucial for maintaining tight timing constraints and preserving computational slack for control algorithms, as discussed in Section III-B.}

\subsubsection{\revrej{Integration and Configuration}}

\revrej{The CLIC is configured with 128 interrupt lines by default, accommodating 64 interrupts from mailboxes and additional interrupts from other peripherals.
To support automatic context saving via the \texttt{fastirq} extension, a \emph{shadow} port is injected \tvlsirevdel{alongside the standard data and instruction ports} through the system interconnect \tvlsirev{from the dual-banked register file}.
As illustrated in~\cref{fig:cpulplet-archi}, this port (highlighted in red) handles the saved context, pushed to memory where the stack resides after register banks are swapped upon interrupt reception.}

The \rev{system} \gls{dma} engine used for \gls{d2d} communication is configured \tvlsirevdel{able,} with \tvlsirevdel{a default}\SI{64}{\bit} data width.
\tvlsirevdel{The exact configuration of the outgoing AXI4 port depends on the interconnect layer between the controller and the controlled system and the requirements of the control application; our configuration serves as a default setup and demands tuning for the specific use case.}
As shown in~\cref{fig:cpulplet-archi}, the \gls{dma} feeds into the bypass network before selecting the I\revdel{/}O protocol of the interface (native AXI4 or \gls{d2d} link), bypassing the traffic on the main system interconnect to reduce unpredictable delays in the sensor readout path.
For off-chip transfers, the IO \gls{dma} is configured with \SI{32}{\bit} data width, matching the sizing of the APB bus.

\subsection{\revrej{\Gls{d2d} link}}\label{subsec:archi:d2d}


\begin{figure}[t]
    \centering
    \includegraphics[width=\columnwidth]{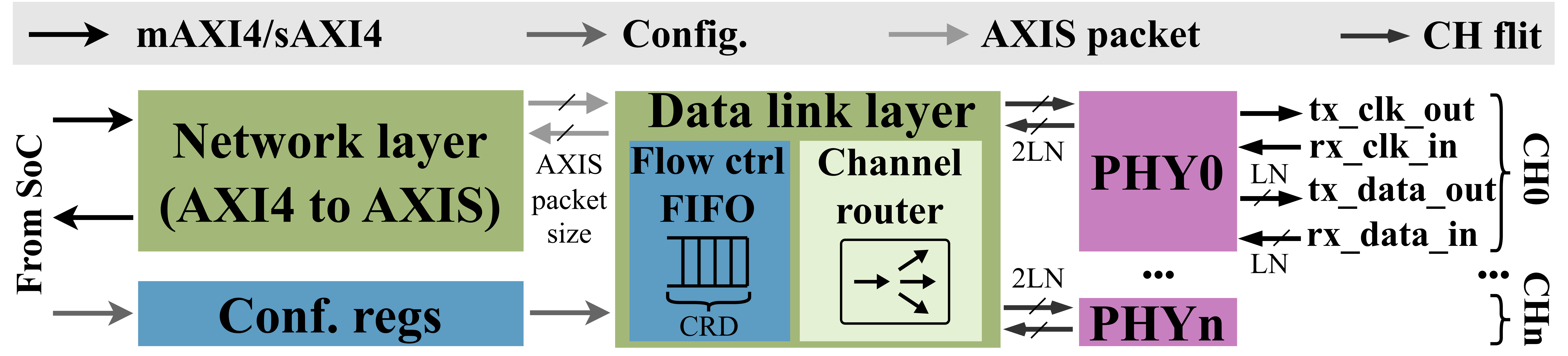}
    \caption{%
        \revrej{\Gls{d2d} link architecture overview.}
    }
    \label{fig:d2d-link-archi}
\end{figure}

\begin{figure}[t]
    \centering
    \includegraphics[width=\columnwidth]{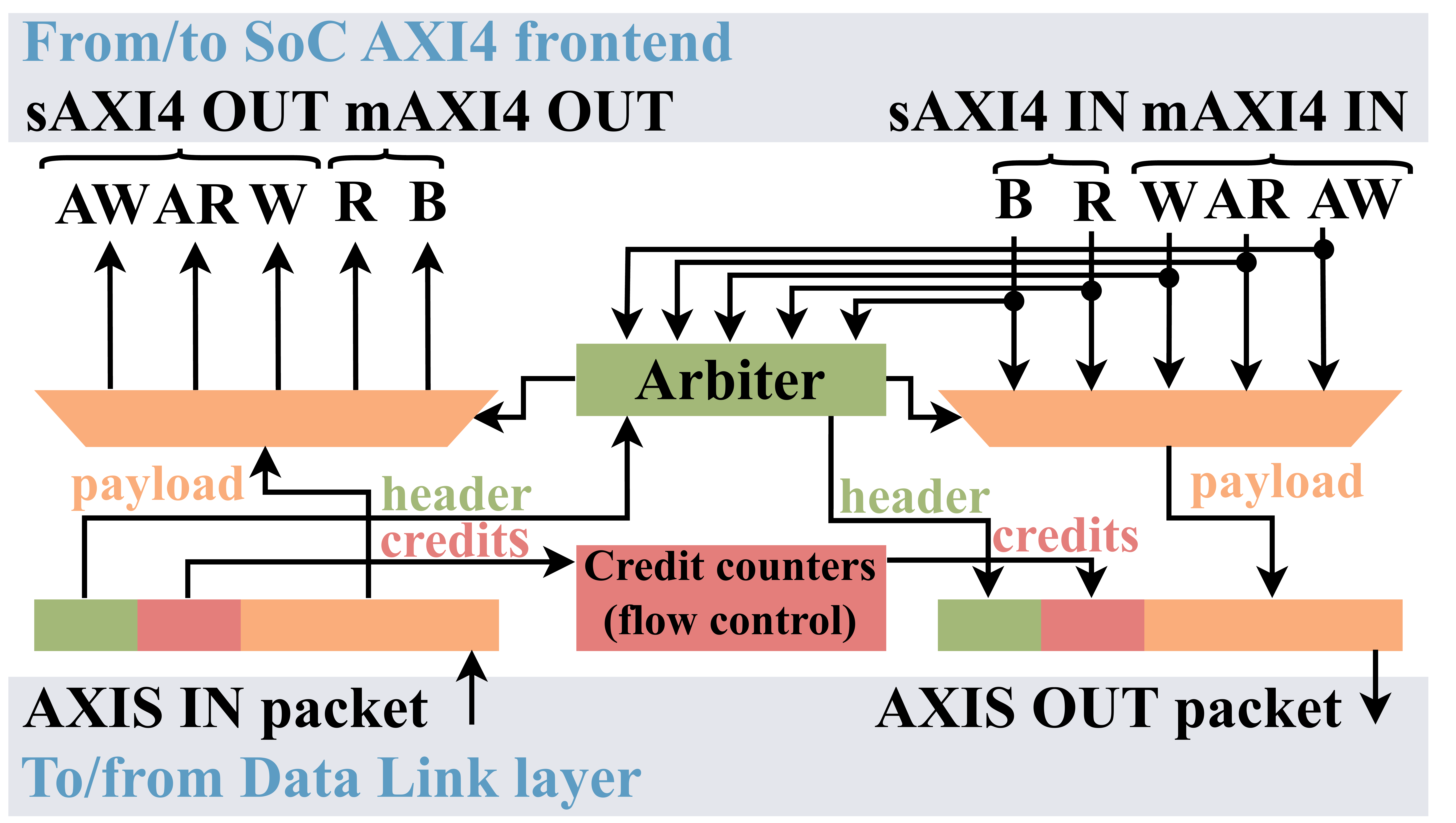}
    \caption{%
        \Gls{d2d} link network layer.
    }
    \label{fig:d2d-link-ntwrk}
\end{figure}

\begin{figure}[t]
    \centering
    \includegraphics[width=\columnwidth]{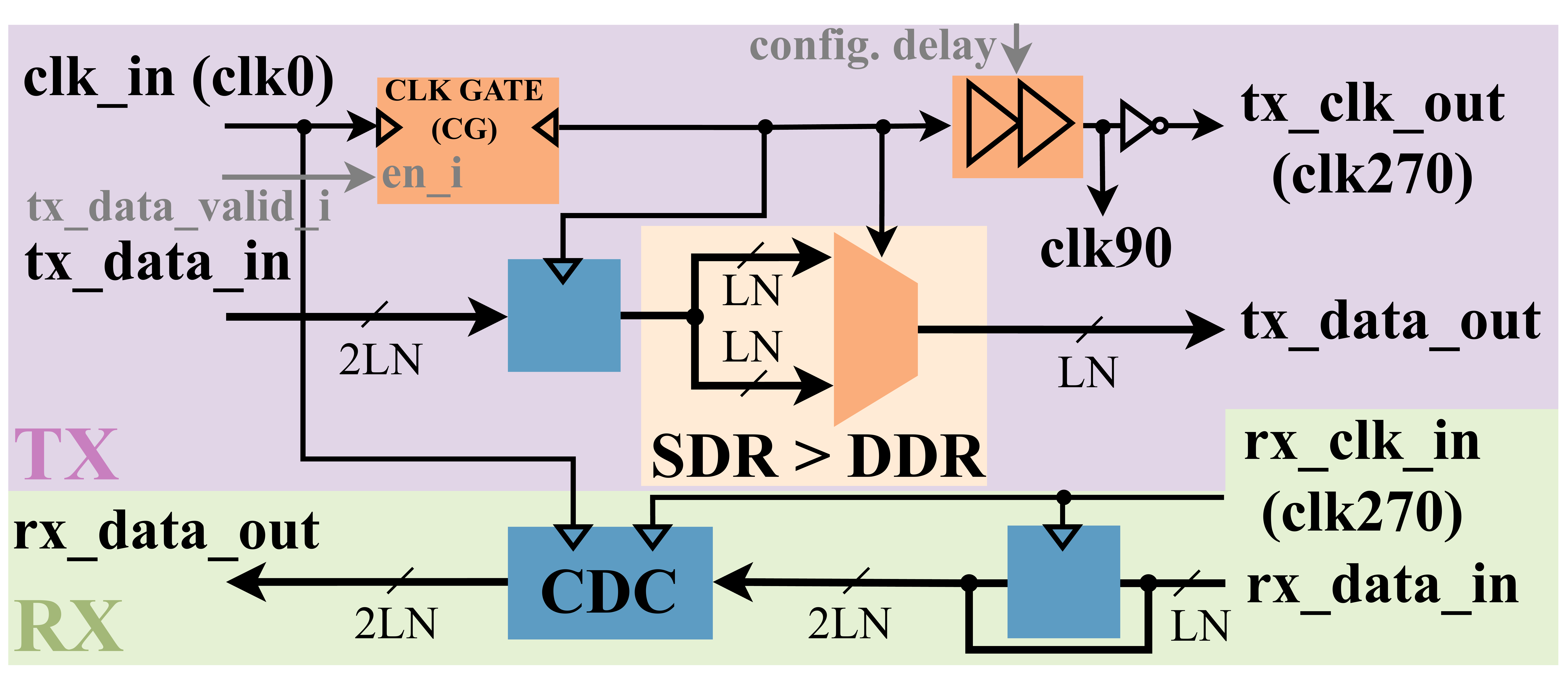}
    \caption{%
        \Gls{d2d} link PHY for one channel. 
        Highlighted in purple and green are the TX and RX modules, respectively, and their main data and clock signals.
    }
    \label{fig:d2d-link-phy}
\end{figure}

The \gls{d2d} link is a digital, source-synchronous link employing \gls{ddr} signaling, designed to translate AXI4 transactions into an off-chiplet interface with a configurable number of channels ($CH$).
The channels operate independently, ensuring isolation and flexibility\tvlsirevdel{ in communication}.
\revrej{A channel is composed of a configurable number of data lanes (\texttt{LN}), with a source-synchronous clock in each direction. 
The channels are synchronized at the receiving end of the link.
The microarchitecture of a \gls{d2d} link is shown in~\cref{fig:d2d-link-archi}. 
It implements \tvlsirevrep{the three lowest layers of the open system interconnection (OSI) reference model}{three layers}: \emph{network}, \emph{data}, and \emph{physical} (PHY) layer.}

\subsubsection{\revrej{Network Layer}}
\revrej{The network layer provides a duplex (manager/subordinate) AXI4 frontend interface\tvlsirevdel{ to ease integration in compatible interconnect fabrics}. 
\tvlsirev{For embedded control chiplets operating on a dedicated control plane, as in our design, AXI4 offers an effective trade-off between simplicity, robustness, and performance, compared to full protocol stacks like PCIe.}
\tvlsirevrep{It}{The network layer} serializes duplex AXI4 requests and responses into an \gls{axis} interface, employing a credit-based flow control mechanism to generate backpressure on the AXI4 frontend.}

\revrej{\Cref{fig:d2d-link-ntwrk} illustrates the network layer circuitry.
The ten channels of the \emph{m\_AXI4} and \emph{s\_AXI4} ports, as labeled in the figure, are arbitrated based on their transfer direction, thus grouping input/output requests and responses.
Flow control is managed through credit counters, which decrement when transactions are issued by the \gls{d2d} (e.g., address write/read or data write transmissions) and increment when transactions are received by the \gls{d2d} (e.g., write responses or read requests).
This credit-based flow control mechanism generates back-pressure on the AXI4 frontend, preventing buffer overflow on the receiver side or guaranteeing no data loss.}

\tvlsirevdel{For each transfer direction, an AXIS \emph{packet} is formed by concatenating: \textbf{(i)} An AXI4 \emph{beat} from one of the five AXI4 channels; the size of the \emph{payload} is determined by taking the largest width among the five AXI4 channels, and reads: max(size(CH\_i) for i in \{\texttt{AW}, \texttt{W}, \texttt{B}, \texttt{AR}, \texttt{R}\}); \textbf{(ii)} the \texttt{B} channel, which is always transmitted to complete write transactions, of width size(\texttt{B}); \textbf{(iii)} a header encoding the AXI4 beat type (4 b); \textbf{(iv)} the number of flow control credits CRD, of width ceil(log2(CRD)) b.}
\tvlsirev{For each transfer direction, the network layer converts AXI4 transactions into AXIS packets, multiplexing the five AXI4 channels (\texttt{AW}, \texttt{W}, \texttt{B}, \texttt{AR}, \texttt{R}) into a unified stream (Fig. 5). Each AXIS packet includes: \textbf{(i)} one AXI beat from the selected channel, with a payload width equal to the maximum of all channel widths; \textbf{(ii)} an optional \texttt{B} channel response (used to complete write transactions); \textbf{(iii)} a 4-bit header tagging the AXI channel type; and \textbf{(iv)} a credit count field to support backpressure. The receiver decodes each packet and dispatches the payloads to the appropriate AXI channel. The dataflow of header, payload, and credits in a packet is shown in Fig. 6.}

\tvlsirevrep{Arbitration between \texttt{AW}/\texttt{AR} and \texttt{W}/\texttt{R} beats}{The arbitration policy prioritizes control channels (\texttt{AR}, \texttt{AW}) and} follows a round-robin policy to prevent starvation and violation of 
AXI4 ordering rules~\cite{arm_amba_2023}.
\tvlsirevdel{Arbiters prioritize \texttt{B} responses as they pertain to \texttt{AW} requests issued from the other manager or subordinate port.}
The design does not support transactions issued but not yet completed (outstanding), allowing only one \texttt{AW}/\texttt{AR} transfer with the same transaction ID at a time.
\tvlsirevdel{Moreover, \texttt{W} requests are not granted before the corresponding \texttt{AW} request.}

\subsubsection{\revrej{Data Link Layer and Flow Control}}
\revrej{The data link layer (\cref{fig:d2d-link-archi}) segments the \gls{axis} payload of an \gls{axis} packet into channel-sized flits.
\tvlsirevrep{If}{When} \gls{ddr} is \tvlsirevrep{supported}{enabled}, each channel has width $2 \cdot \texttt{LN}$, \tvlsirev{as shown in Fig. 4 on the right of the Data link layer module,} for data transmission and reception\tvlsirev{, which thereby happens in one clock cycle.} 
This does not account for the transmitting and receiving clocks.
In this setting, the number of bits transferable per clock cycle, or the theoretical bandwidth, is expressed as $\theta = 2 \cdot CH \cdot LN$ \si{b}.}

\revrej{The data link layer includes a flow control credit buffer to store incoming data.
This buffer operates as a FIFO (First-In, First-Out) queue. In a given \gls{d2d} instance, data is pushed into the FIFO during reception and popped out during transmission. 
A similar process occurs symmetrically in the FIFO queue on the other \gls{d2d} instance. 
The network layer's flow control mechanism ensures that the FIFO does not overflow by applying backpressure as needed.
To prevent deadlocks during long write/read bursts, e.g. from the system \gls{dma}, the network layer can send empty payload packets with credits to match the capacity of the burst transfer.}

\revrej{Each FIFO element has a size of $\theta$, meaning it contains all the data flits across all channels.
The FIFO depth equals the number of flow control credits $CRD$ times the number of splits (or chunks) required to divide a packet into smaller segments: $CRD$ $\cdot$ (size(packet) / $\theta$).
One should aim at configuring the link to preserve the original packet size, minimizing the number of chunks, else increasing the FIFO depth.
The $CRD$ parameter plays a critical role in determining the link throughput, particularly for large write/read bursts where the FIFO dimensions constrain the backpressure efficiency. 
This impact is analyzed in detail in \cref{subsec:eval:bw_lat}.}

\revrej{A channel router, shown in~\cref{fig:d2d-link-archi}, combinationally splits the stream in flits and distributes them across the channels. When \tvlsirev{$CH=1$}, the router module is bypassed in hardware to save area.}
\tvlsirev{Faulty channel support could be incorporated into the D2D design by extending the data link layer with hardware-level detection and suppression during link calibration. We leave this enhancement to future work.}

\begin{figure}[t]
     \centering
     \includegraphics[width=\columnwidth]{fig/d2d_chronogram.pdf}
     \caption{%
         \revrej{\Gls{d2d} chronogram of an exemplary \SI{32}{\byte}-burst \gls{dma} write transfer through the \gls{d2d} link.
         We show the network, link, and PHY TX signal interfaces of the transmitting \gls{d2d} link instance and the PHY RX signal interface of the receiving \gls{d2d} link instance.
         Only one channel is shown for readability.}
     }
     \label{fig:d2d-link-chronograms}
 \end{figure}

\subsubsection{\revrej{PHY Layer}}
The PHY layer of a single channel provides access to the data lanes and the transmitting/receiving clock signals across all channels. Each channel includes both a transmission (TX) module and a receiver (RX) module\tvlsirev{, highlighted in Fig. 6 in purple and green, respectively.}
When \gls{ddr} signaling is \tvlsirevrep{employed}{enabled}, the channel flit \tvlsirev{\texttt{tx\_dat\_in}, of size $2 \cdot \texttt{LN}$,} is transmitted \tvlsirevrep{on two consecutive clock edges, as depicted in the figure}{in one clock cycle}. \tvlsirevrep{This configuration ensures alignment between the channel flit size ($2 \cdot \texttt{LN}$) at the data link layer and the data interface width of the D2D link, i.e., \texttt{LN}, as illustrated in the diagram.}{This means that the data interface exposed by the link, of size \texttt{LN} on the transmission and reception paths, sends or receives data every half clock cycle to keep up with the bandwidth supplied by the data link layer flits.}
\tvlsirev{At the circuit level, on the TX path, we use a clock multiplexer to alternate the transmission of $LN$ data bits depending on the clock signal's high or low phase, thereby achieving DDR signaling.}
Alternatively, \gls{sdr} signaling can be used \tvlsirev{bypassing the clock multiplexing logic}, albeit with a potential reduction in performance.\tvlsirevdel{, unless mitigated by increasing the PHY clock frequency while maintaining the same data lane size (\texttt{LN}) or doubling the latter.}

\revrej{The TX module forwards a source-synchronous clock, \emph{tx\_clk\_out}, off-chiplet alongside the channel flit.
This forwarded clock incorporates a three-quadrature (270\textdegree) phase shift relative to the PHY input clock, \emph{clk\_in}, to maximize the data sampling window. 
The phase shift is achieved using a fully digital, configurable delay line implemented as a binary tree of multiplexers. This structure shifts the clock by 90\textdegree, and a cascade inverter is used to achieve the final three-quadrature phase shift, illustrated in \cref{fig:d2d-link-phy}.
This three-quadrature shifting optimizes the data sampling eye window, enabling robust data transmission.
To ensure data is sampled only when valid, the TX module employs a \gls{cg} cell. The \gls{cg} is controlled by the \emph{tx\_data\_valid\_i} signal and modulates the logic level of the PHY's source clock prior to phase shifting.
The data-valid/ready handshake signals are generated at the data link layer and are derived from their corresponding signals in the AXI4 and AXIS frontends injected at the network layer.
On the RX side, data sampling is performed using the received clock, \emph{rx\_clk\_in}. 
A \rev{FIFO-based} \gls{cdc} with 2-stage flip-flop synchronizers is responsible for synchronizing the incoming data stream, \emph{rx\_data\_in}, with the PHY source clock \emph{clk\_in}.
\tvlsirev{At the circuit level, when DDR is enabled, the \texttt{rx\_data\_in} input of size \texttt{LN} is sampled every cycle on the falling edge of the receiving clock (thus, every half-clock cycle), as illustrated by the flip-flop in the bottom-right corner of Fig.~6. This value is then coalesced with the sample captured on the rising edge of ControlPULP’s clock to form a $2 \cdot LN$-bit data flit over a full clock cycle, fed to the CDC described above.}
The total number of wires exposed by the PHY is given by $N_{wrs} = CH \cdot (2 \cdot (LN + 1))$, which accounts for the data lanes and the forwarded clock in both input and output directions (duplex).}

\subsubsection{\revrej{Latency}}
\revrej{\Cref{fig:d2d-link-chronograms} illustrates the chronogram of a \SI{32}{\byte} burst transfer initiated by the system \gls{dma} via a \gls{d2d} link configured with \tvlsirev{$LN=\SI{8}{\bit}$} and \tvlsirev{$CH=8$}. 
The diagram showcases the network, link, and TX PHY interfaces of the transmitting \gls{d2d} instance, along with the RX PHY interface of the receiving instance. 
Only the first channel of the data link layer and PHY is depicted. The transmitting and receiving instances are interconnected to form the inter-chiplet bridge.}

\revrej{At the network layer (yellow), a \SI{4}{\byte} address is first transmitted on the \texttt{AW} channel, followed by the \SI{32}{\byte} data on the \texttt{W} channel. Since the system \gls{dma} has a \SI{64}{\bit} data width, \SI{8}{\byte} are dispatched per clock cycle. The handshake signals associated with the transfer are also shown. 
The number of clock cycles between the address and data transmissions depends on the implementation of the \gls{dma} and the controller interconnect. For simplicity, this delay is depicted as 1 clock cycle, denoted by the symbol $\approx$.
The AXI4 interface is then converted into an AXIS interface, represented by the tuple \emph{tx\_axis\_out\_req\_o.t} and \emph{tx\_axis\_out\_req\_o.t\_valid}.
The latency for this conversion, $T_{\text{AXI4-AXIS}}$, is at most 1 clock cycle.
The \gls{d2d} link configuration ensures that the AXIS data payload size fully encapsulates the upstream AXI4 transactions.}

\revrej{At the data link layer (red), the \emph{tx\_data\_out} payload for a single channel is shown, with each flit being $2 \cdot \texttt{LN} = \SI{2}{\byte}$. 
The handshake interface for the data link layer is also included. 
The channel flit splitting of the AXIS stream and its subsequent routing incur no additional latency.}

\revrej{The PHY layer illustrates the TX interface of the transmitting \gls{d2d} link (purple) and the RX interface of the receiving \gls{d2d} link (green) for one channel. 
At the PHY TX layer, the \emph{tx\_data\_out} signal is transmitted on both edges of the transmitted clock (\emph{tx\_clk\_out}), resulting in a data rate of \SI{1}{\byte} per half clock cycle in this example.
The PHY TX layer incurs a latency, $T_{\text{RX,CDC}}$, of 1 clock cycle due to the clock gating cell, which latches the enable signal at the rising edge of \emph{clk\_in} and applies it at the subsequent clock cycle.}

\revrej{Once the data and clock signals leave the PHY TX of the transmitting \gls{d2d} instance, they are synchronized by the RX module of the receiving \gls{d2d} instance. 
The inter-chiplet latency between the two links is determined by the IO, interposer, and packaging solution. 
This delay is depicted as $T_{\delta}$ in the chronogram and is set arbitrarily to 1 clock cycle for readability.
On the receiving PHY (RX PHY), up to three additional clock cycles are required to synchronize the \emph{rx\_data\_in} signal with \emph{rx\_data\_out} between \emph{rx\_clk\_in} and \emph{clk\_in} via the \gls{cdc}.
This delay is \revrep{attributed}{due} to the synchronization and buffer logic within the \gls{cdc}.}

\revrej{After the PHY RX layer in the receiving \gls{d2d} instance, the data path (data link and network layer) mirrors that of the transmitter and thus is omitted from \cref{fig:d2d-link-chronograms}.}

\tvlsirev{While the \gls{d2d} link is not fully latency-insensitive in the theoretical sense, since it relies on finite buffering and timely credit returns, it achieves a high degree of latency tolerance through handshake-based elastic buffering and credit-based flow control. This design absorbs link-level jitter and moderate inter-chiplet delays, as we show in~\cref{subsec:eval:bw_lat}.}

\section{Experimental Results}\label{sec:eval}

\begin{figure}[t]
    \centering
    \includegraphics[width=\columnwidth]{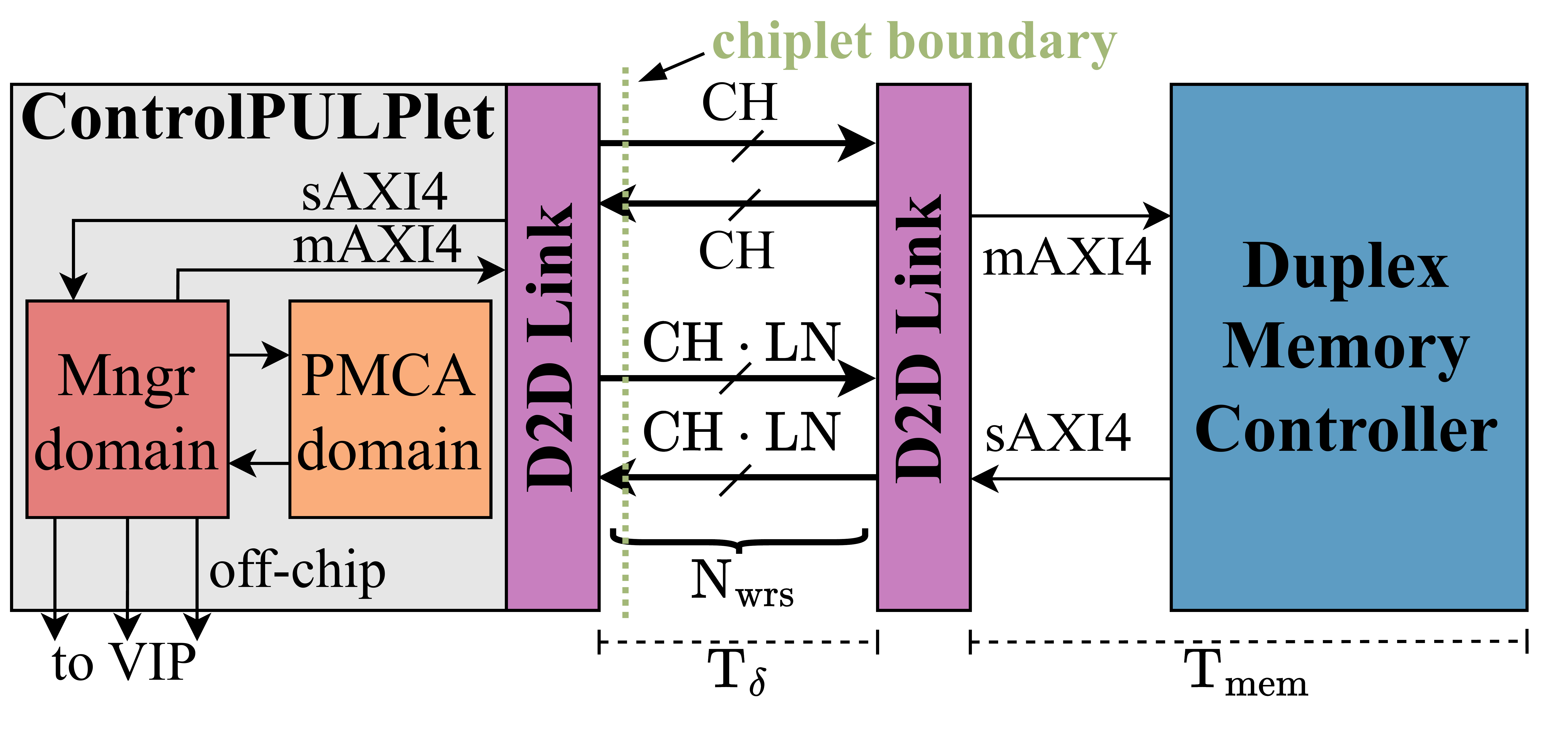}
    \caption{%
        \revrej{Simulation setup.
        The \gls{d2d} interface of a ControlPULPlet instance is connected to another \gls{d2d} instance and then to an AXI4 duplex memory controller. The testbench environment uses the \gls{rtl} description of the actual hardware; it is, therefore, cycle-accurate.}
    }
    \label{fig:eval:cpulplet-sim-loopback}
\end{figure}

\begin{figure}[t]
    \centering
    \includegraphics[width=\columnwidth]{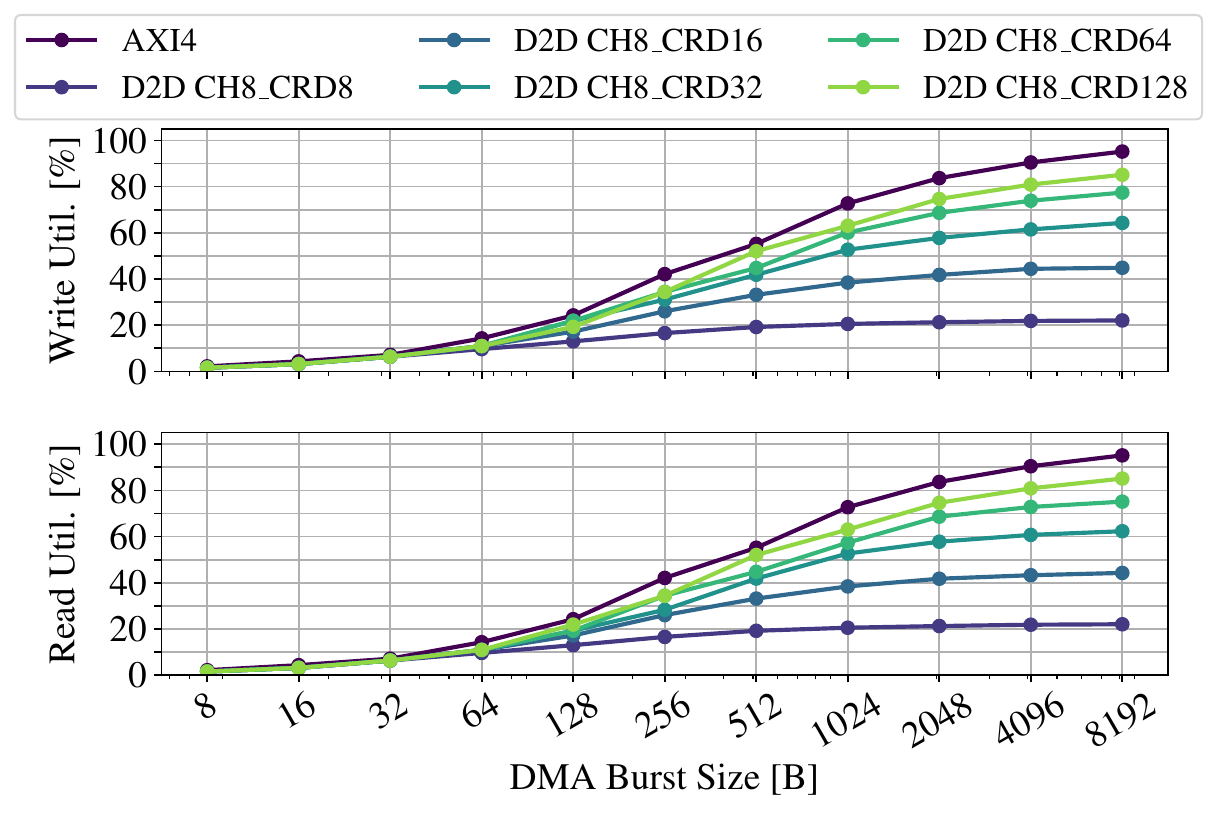}
    \caption{%
        Throughput comparison between \gls{d2d} link and AXI4 with \tvlsirev{$CH=8$} and \tvlsirev{$LN=\SI{8}{\bit}$} at varying $CRD$ and system \gls{dma} burst size in bytes.
    }
    \label{fig:eval:cpulplet-bus-util}
\end{figure}

We first \revrep{perform}{present} a functional evaluation of the \gls{d2d} throughput (\cref{subsec:eval:bw_lat}). 
Then, we assess and discuss \tvlsirevrep{its impact}{the impact of the \gls{d2d} interface and real-time features} on an exemplary \gls{dvfs} control loop case study (\cref{subsec:eval:case_study}).
\revrej{\Cref{fig:eval:cpulplet-sim-loopback} illustrates the testbench setup for cycle-accurate \gls{rtl} simulations, which comprises ControlPULPlet and its \gls{d2d} link instance, and a destination \gls{d2d} link instance. The vertical dashed green line shows the chiplet boundary. 
The destination \gls{d2d} link connects to a duplex downstream memory controller~\footnote{Available open-source at \url{https://github.com/pulp-platform/axi}}.
This choice is preferred over a loopback configuration, which would connect ControlPULPlet's \gls{d2d} link’s input and output, as it allows modeling the storage elements of sensors/actuators for the controller policy.
\tvlsirevdel{These storage elements, usually registers, vary by application domain, particularly their physical floorplan and the address map (linear or with stride).}
The latency of the interconnect and memory system after the receiving \gls{d2d} instance is represented as $T_{\text{mem}}$. 
In this simulation setup, both $T_{\delta}$ and $T_{\text{mem}}$ are configurable.
\Cref{fig:eval:cpulplet-sim-loopback} also highlights ControlPULPlet's off-chip interface, connected to verification IPs (VIPs)\tvlsirevdel{, such as EEPROM and Flash memories,} to simulate communication along this path.}

\subsection{\Gls{d2d} Throughput}\label{subsec:eval:bw_lat}

\subsubsection{Sizing the \gls{d2d} Link}
Finding an optimal \gls{d2d} link configuration requires determining the minimum values of $LN$ and $CH$ that can sustain the bandwidth of the native AXI4 interface after serialization into the AXIS protocol, namely $\theta \geq$ AXIS packet size.
As described in \cref{subsec:archi:d2d}, the size of an AXIS packet includes a constant and variable component.
The constant component is determined by the system-level configuration of the five AXI channels. 
It is approximately $\SI{100}{\bit}$ for a configuration with $\SI{64}{\bit}$ data and $\SI{32}{\bit}$ address widths, as for ControlPULPlet's system \gls{dma}.
The variable component depends on the depth of the control flow FIFO, as credits are sent/received over the \gls{d2d} interface.
When using the native AXI4 interface, the \gls{dma} can transfer up to \SI{64}{\bit} per clock cycle, i.e. $\theta = \SI{64}{\bit}$.
To ensure that the \gls{d2d} link matches this theoretical bandwidth, a channel width of $CH \geq 7$ and $LN \geq 8$ \si{\bit} \tvlsirevrep{guarantee}{guarantees} that $\theta \geq \SI{112}{\bit}$ can be transferred at each clock cycle.
For this evaluation, we select an upper bound \tvlsirev{$CH=8$} and \tvlsirev{$LN=\SI{8}{\bit}$}. This makes the flow control FIFO size dependent on the choice of $CRD$ only.
Additionally, we configure $T_{\delta}=0$ and $T_{\text{mem}}=1$ clock cycles to minimize external overhead in the throughput evaluation.

\subsubsection{Bus Throughput and Utilization}
Once the theoretical bandwidth $\theta$ is determined, the maximum attainable duplex throughput of the \gls{d2d} link reads $\Theta = 2 \cdot \alpha \cdot f \cdot \theta$ \si{\bit\per\second}, where $\alpha \in [0,1]$ is the relative \emph{bus utilization}, and $f$ is the operating frequency.
At $f = \SI{200}{\mega\hertz}$, this configuration achieves a theoretical bandwidth of $\Theta = \alpha \cdot$ \SI{51.2}{\giga\bit\per\second}. 
It represents the bus inefficiencies, i.e., clock cycles wasted without transferring data.
In the context of our \gls{d2d} design, $\alpha$ is influenced by factors including the additional conversion latencies introduced by the AXI4 frontend (\cref{fig:d2d-link-chronograms}) and the limited storage capacity of the flow control FIFO.
Consequently, even when $\theta$ is carefully selected to match the theoretical bandwidth of the native on-chip AXI4 interface, $\alpha$ inherently reflects the impact of the architectural design choices made for the \gls{d2d} link, affecting the overall throughput $\Theta$.

\subsubsection{Results and Discussion}
We measure the \gls{d2d} link's read and write paths throughput $\Theta$ at varying $CRD$. 
We compare it with the throughput achieved by the native AXI4 on-chip interface without \gls{d2d} link.
For this purpose, the system \gls{dma} is programmed to issue repeated write and read transfers at increasing burst sizes, starting from \SI{8}{\byte}.

Results are shown in \cref{fig:eval:cpulplet-bus-util}.
The native on-chip interface, labeled as \emph{AXI4} and represented in purple, approaches peak utilization ($\alpha=1$) starting from burst sizes of \SI{2}{\kibi\byte}, the maximum permitted by the AXI4 specifications.
The \gls{d2d} interface achieves approximately \SI{85}{\percent} bus saturation for large bursts on the write and read paths with $CRD=128$, compared to \SI{95}{\percent} of the native AXI4 on-chip performance. 
\tvlsirevdel{This accounts for at most 90\% the bandwidth requirement.} 
The performance gap can be narrowed by increasing further the FIFO depth $CRD$, but results in greater area overhead for the \gls{d2d} link and may necessitate a different implementation strategy for the flow control buffer, e.g., an SRAM-based design.
This area cost is evaluated in-system in~\cref{subsec:eval:silicon_perf}.

\subsection{\revrej{Case study: Online \glsentrytext{dvfs} Control Loop of \glsentrytext{hpc} Processors}}\label{subsec:eval:case_study}


\begin{figure}[t]
    \centering
    \includegraphics[width=\columnwidth]{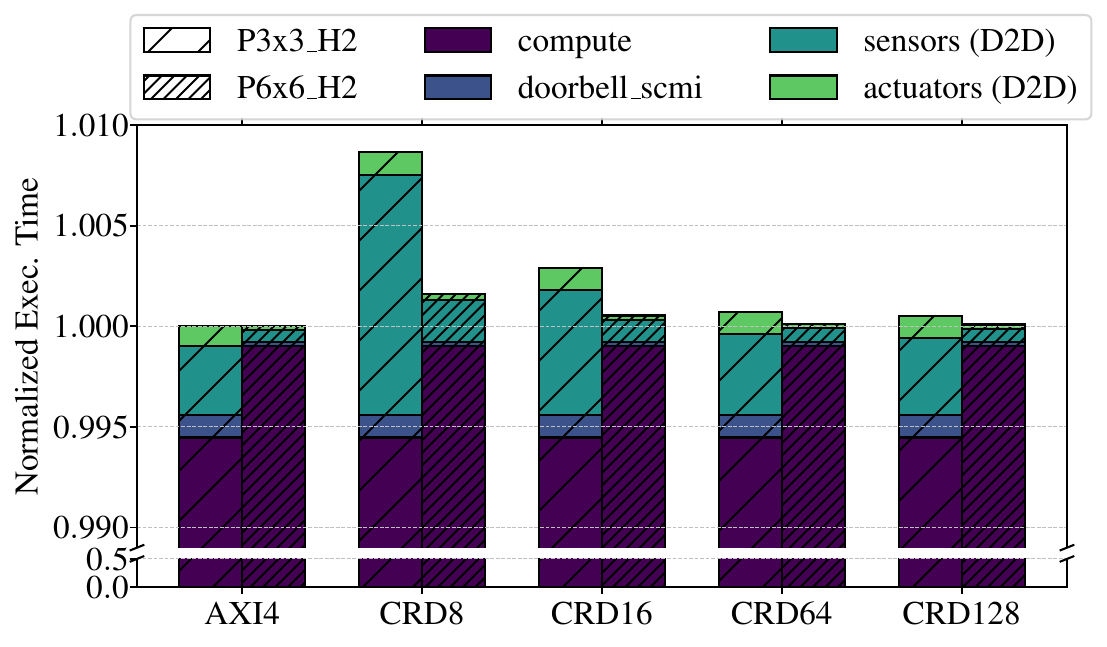}
    \caption{%
        \revrej{Impact of \gls{d2d} communication over native on-chip communication (AXI4) on a \gls{dvfs} CPU control scenario at a varying number of \tvlsirev{$CRD$}. \tvlsirev{$CH=8$} and \tvlsirev{$LN=\SI{8}{\bit}$} for the \gls{d2d} link.} \todo{TODO: replace scmi with irq}
    }
    \label{fig:eval:cpulplet-case-study-d2d}
\end{figure}

We consider a control scenario involving online \gls{dvfs} of \gls{hpc} processors.
In this application, the controlled system includes $N_c$ application-class \glspl{pe}\tvlsirevdel{ arranged in a grid}. 
The sensory layer consists of on-chip \gls{pvt} sensors\tvlsirevdel{ monitoring static and dynamic chip metrics and off-chip monitoring of voltage rails}. 
The actuation layer includes setting on-chip \glspl{pll} for per-core frequency scaling and off-chip voltage regulators for voltage scaling. \tvlsirevdel{Voltage scaling occurs per domain, with multiple PEs sharing a common voltage level.}
\tvlsirevdel{Sensing and actuation of on-chip elements happen through ControlPULPlet's D2D link or on-chip AXI4 interface, depending on the configuration.} 
We consider 500 \gls{pvt} sensor registers of \SI{8}{\byte} each. Similarly, we assume one \gls{pll} per-core\rev{.}
\tvlsirevdel{Off-chip accesses are performed via the IO DMA-regulated off-chip interface.}
Setpoints, such as the reference \gls{dvfs} operating point \tvlsirevdel{(frequency-voltage tuple)}, are asynchronously sent to the controller via the mailbox unit\tvlsirevdel{ using the SCMI protocol}.
\tvlsirevdel{The operating system on top of the PEs typically dispatches these directives, acting as a high-level controller.}
The online control policy solves \tvlsirevrep{an}{a floating-point} \gls{mpc} problem \tvlsirevrep{ using the SoA OSQP solver library, which applies the ADMM to iteratively solve a linear system and}{periodically, which} outputs the controlled \gls{dvfs} points\tvlsirevdel{through their respective interfaces}. \tvlsirev{The MPC uses a prediction horizon $H_p = 2$.}
\tvlsirevdel{The thermal and power model for the control is based on the finite element spatial discretization of the partial differential equations (PDEs) governing the silicon and copper heat dissipation.}For a detailed discussion \tvlsirevrep{of this model}{of the MPC power and thermal model}, we refer the reader to~\cite{bambini2024arxiv}.
\tvlsirevdel{The MPC problem size scales with the number of controlled PEs.} 
We define the \gls{dvfs} control problem for each \gls{pe} \tvlsirevdel{with a prediction horizon $H_p$} as $\mathtt{P N_c \times N_c\_{Hp}}$.

\subsubsection{\revrej{Evaluation Settings}}
We configure the \gls{rtl} testbench setup in~\cref{fig:eval:cpulplet-sim-loopback} with $T_{\delta}=50$ and $T_{\text{mem}}=100$ clock cycles. 
We consider this a reasonable upper bound for \revdel{deep} NoC-based interconnect systems with single-cycle access to on-chip memory\tvlsirevdel{ or registers}~\cite{Ottaviano2024}.
Two synchronous, periodic tasks are initiated by FreeRTOS using the platform's timers: a task for voltage sensing and actuation through the off-chip interface and a task for temperature sensing, control policy computation, and frequency actuation through the \gls{d2d} interface.
The two tasks have different \tvlsirevrep{speeds, as voltage can change on the $\mu s$ scale, while temperature variations lie in the ms range. We select }{periods:} $T_{\text{short}}=\SI{250}{\micro\second}$ and $T_{\text{long}}=\SI{3}{\milli\second}$, respectively\tvlsirevrep{, with a controller frequency of}{. ControlPULPlet runs at} \SI{500}{\mega\hertz}.
\tvlsirevdel{For every MPC step,}The control policy computation is handled by the manager \tvlsirevdel{domain} core and terminated after 5 iterations\tvlsirevdel{; it is carried out with 32 b floating-point precision}; the \gls{pmca} is not used, as \gls{mpc} acceleration is outside this work's scope. 
\tvlsirevdel{The system DMA handles PVT sensor readouts via the D2D link; actuator dispatch remains on the manager core for inter-chiplets transfers. For off-chip transfers, sensor and actuator communication is handled by the IO DMA.} 
\tvlsirevrep{SCMI}{M}essage dispatching is simulated by issuing messages \tvlsirevdel{from the testbench}to ControlPULPlet’s mailbox at the beginning of each period with the target \gls{dvfs} setpoint.

\tvlsirev{Overall, meeting the stringent real-time constraints of this application requires close HW/SW co-design.}
\ifx\showtvlsirebuttal\undefined
\else
\subsubsection{\sout{Performance Metrics}}
\fi
We define \emph{slack} as the free time available to the controller before the \tvlsirevrep{current hyperperiod}{task with the longest period} ends\tvlsirevdel{and the next begins}. Denoting the computation and communication times required by the controller as $T_{\text{ctrl}} = T_{\text{comp}} + T_{\text{comm}}$, the slack is calculated as
$T_{\text{slack}} = T_{\text{long}} - T_{\text{ctrl}}$.
To meet the deadline, \tvlsirevrep{all computation and communication must be completed within the control hyperperiod, ensuring}{we must always ensure} $T_{\text{slack}} > 0$.
\tvlsirevdel{Slack can be increased by extending the hyperperiod or reducing $T_{\text{ctrl}}$. Application requirements and physical limitations typically constrain the former, while the latter can be optimized through hardware/software co-design of critical controller components.}
\tvlsirevdel{Our evaluation demonstrates how ControlPULPlet's hardware extensions, described in Sec. III-B, maximize the slack with negligible performance penalty from the D2D link.}

\subsubsection{\revrej{Results and Discussion}}

We first highlight the benefits of fast interrupt handling and autonomous \gls{dma} before focusing on \gls{d2d} link communication.
The time to decode a message \tvlsirev{dispatched to the mailbox unit} is given by: 
\tvlsirev{
\[
T_{\mathrm{MSG}}
= T_{\mathrm{CLIC\text{-}to\text{-}ISR}}
+ T_{\mathrm{ISR}}
+ T_{\mathrm{ISR\text{-}to\text{-}dec}}
\]
}
\tvlsirevrep{where $T_{\text{CLIC-to-ISR}}$ is the time from when the CLIC interrupt controller accepts a doorbell interrupt to the first instruction in the ISR, $T_{\text{ISR}}$ is the time to execute the ISR, and $T_{\text{ISR-to-dec}}$ accounts for deferring the message decoding task, which consumes $T_{\text{dec}}$. Decoding is deferred to keep $T_{\text{ISR}}$ as short as possible.}{where $T_{\text{CLIC-to-ISR}}$ is the time from CLIC acceptance to the first \gls{isr} instruction, $T_{\text{ISR}}$ is the ISR execution time, and $T_{\text{ISR-to-dec}}$ is the deferred decoding time, kept outside the ISR to reduce its duration, a common real-time programming practice.}
\tvlsirevdel{Figure 10 demonstrates the improvement in SCMI decoding time due to the \texttt{fastirq} extension.}
\tvlsirevdel{Specifically, }$T_{\text{CLIC-to-ISR}}$ achieves a \SI{72}{\percent} speedup \tvlsirev{(46 to 13 clock cycles)} thanks to automatic hardware-based context saving and restore \tvlsirevrep{.}{and, }reduces the decoding time by \SI{13}{\percent}. This acceleration is critical when multiple concurrent \tvlsirevdel{SCMI}{message} requests from various \glspl{pe} \tvlsirevrep{are sequentially}{must be processed by the manager domain}\tvlsirevdel{, as in per-core frequency control scenarios}\tvlsirev{. As described in Sec. II-B, this situation benefits from tail-chaining optimizations in the CLIC, dramatically reducing interrupt handling time to 8 clock cycles.}

\tvlsirevdel{The main advantage of the real-time extension designed for the DMAs is to hide the latency associated with launching and executing periodic data transfers.} 
For \rev{the} system \gls{dma}, programming incurs a fixed cost of $T_{\text{single,DMAprog}} = 100$ clock cycles, \tvlsirev{as it happens once during $T_{\text{long}}$}. During long inter-chiplet transfers, \tvlsirevrep{this offloading}{the automation of this phase} provides significant performance gains.
For \rev{the} IO \gls{dma}, the benefit is even greater. Programming must occur periodically ($T_{\text{short}}$) \tvlsirevdel{while the core executes a task with a longer periodicity ($T_{\text{long}}$). Each IO DMA programming} \tvlsirev{and} requires a context switch\tvlsirevrep{ of the core, adding an overhead of $T_{\text{ctx\_switch}}=100$ clock cycles.}{ by the running RTOS.} 
\tvlsirevdel{In a vanilla controller, the total programming overhead is}: 
\tvlsirev{
\[
T_{\mathrm{tot,DMAprog}}
= \left(\nicefrac{T_{\mathrm{long}}}{T_{\mathrm{short}}}\right)
  \cdot \bigl(T_{\mathrm{single,DMAprog}} + T_{\mathrm{ctx\_switch}}\bigr)
\]
}
With RT-DMA, context switches are eliminated, reducing the overhead to $T_{\text{single,DMAprog}}$. 
For $T_{\text{short}} = \SI{250}{\micro\second}$ and $T_{\text{long}} = \SI{3}{\milli\second}$, this results in a \SI{98}{\percent} speedup\tvlsirevdel{, effectively hiding IO DMA overhead and enabling simultaneous execution with the main core}.
These optimizations \tvlsirevdel{significantly} mitigate $T_{\text{comm}}$ and enhance deadline completion guarantees.


\tvlsirev{Finally,} \cref{fig:eval:cpulplet-case-study-d2d} illustrates the impact of the \gls{d2d} link on control loop performance for two \revrep{scenarios}{control problems} ($\mathtt{P3\times3\_{H2}}$ and $\mathtt{P6\times6\_{H2}}$), 
\tvlsirev{where \texttt{fastirq} and autonomous DMA are always active.}
The execution time is normalized to the native AXI4 \revdel{-based} \rev{baseline}.
For both problems, the baseline already meets the deadline, \tvlsirevrep{and}{with} \tvlsirev{$T_{\text{slack}} = \SI{86}{\percent}$} and \tvlsirev{$T_{\text{slack}} = \SI{21}{\percent}$} \tvlsirevdel{of the hyperperiod}for the smaller and larger problems, respectively.
The \gls{d2d} link is configured with \tvlsirev{$CH=8$}, \tvlsirev{$LN=8$}, and varying flow control FIFO sizes. 
\tvlsirevdel{Using \texttt{fastirq} and autonomous DMA, w}We \revrep{analyze}{report} on message decoding, sensor/actuator communication, and control policy computation.
Compared to the \revdel{native AXI4 interface} baseline, the \gls{d2d} interface \revrep{degrades}{has predictably a higher} $T_{\text{comm}}$. 
For \tvlsirev{$CRD=8$}, latency increases by \SI{3.1}{\times} and \SI{3.2}{\times} for $\mathtt{P3\times3\_{H2}}$ and $\mathtt{P6\times6\_{H2}}$, respectively\revrep{;}{.} \rev{A more efficient D2D configuration. i.e.,} \tvlsirev{$CRD=128$}\rev{,} reduces this \rev{gap} to \SI{1.1}{\times} in both applications, hereby approaching the native AXI4 performance of on-chip control.
\tvlsirevdel{The benefit of an optimally configured D2D is more visible for large transfers that can saturate the bus, like sensor readouts (4KiB bursts), compared to narrow writes handled by the core for actuator control (hundreds of bytes), that have naturally poor bus utilization.}
\revdel{We analyze the impact of $T_{\text{comp}}$ on execution time degradation caused by the D2D link.} 
As shown in~\cref{fig:eval:cpulplet-case-study-d2d}, \rev{in the analyzed application} the compute phase dominates $T_{\text{ctrl}}$, accounting for over \SI{99}{\percent} of execution time. 
Consequently, the \revrep{impact of the D2D link degradation}{D2D link penalty} \rev{on the overall $T_{\text{ctrl}}$} \tvlsirevrep{remains}{is always} negligible, \tvlsirevdel{regardless of its configuration}\rev{i.e., $<$\SI{0.9}{\percent} for \tvlsirev{$CRD=8$} and $<$\SI{0.05}{\percent} for \tvlsirev{$CRD=128$}}.
\tvlsirevdel{This is primarily due to two factors. First, the computational complexity of the MPC algorithm scales with the number of PEs and the ADMM iterations required for convergence, as opposed to the sensor readout phase which is constant.} 
\tvlsirevdel{This is because in our execution all data required for the control policy computation is kept in the manager domain's local SPM, ensuring controller independence, reducing latency, and isolating it from other system components.} 
\tvlsirev{Certainly, i}f the \tvlsirevrep{application}{MPC computation} required fetching \tvlsirevrep{additional}{some of the} data \tvlsirevdel{to perform the computation} from an \tvlsirevrep{external}{off-chip} memory \tvlsirevdel{located on a far-end chiplet}, the memory traffic over the D2D link would increase, as well as \tvlsirevrep{the}{its} penalty\tvlsirevdel{ on $T_{\text{comm}}$}.

\tvlsirev{This section shows that the RT-DMA and an optimally designed D2D link enable high-throughput communication without processor intervention, on par with the native on-chip performance. Fast interrupt handling similarly reduces the processor's response latency and context-switching overhead, allowing it to focus on computation and PMCA offload, if enabled.}
In the next section, we show that the \gls{d2d}-based control also comes at a \revrep{negligible}{minimal} area and energy cost.

\subsection{Silicon demonstrator}\label{subsec:eval:demonstrator}

\begin{figure}[t]
    \centering
    
    \begin{minipage}[t]{\columnwidth}
        \subfloat[]{
            \includegraphics[width=.97\columnwidth]{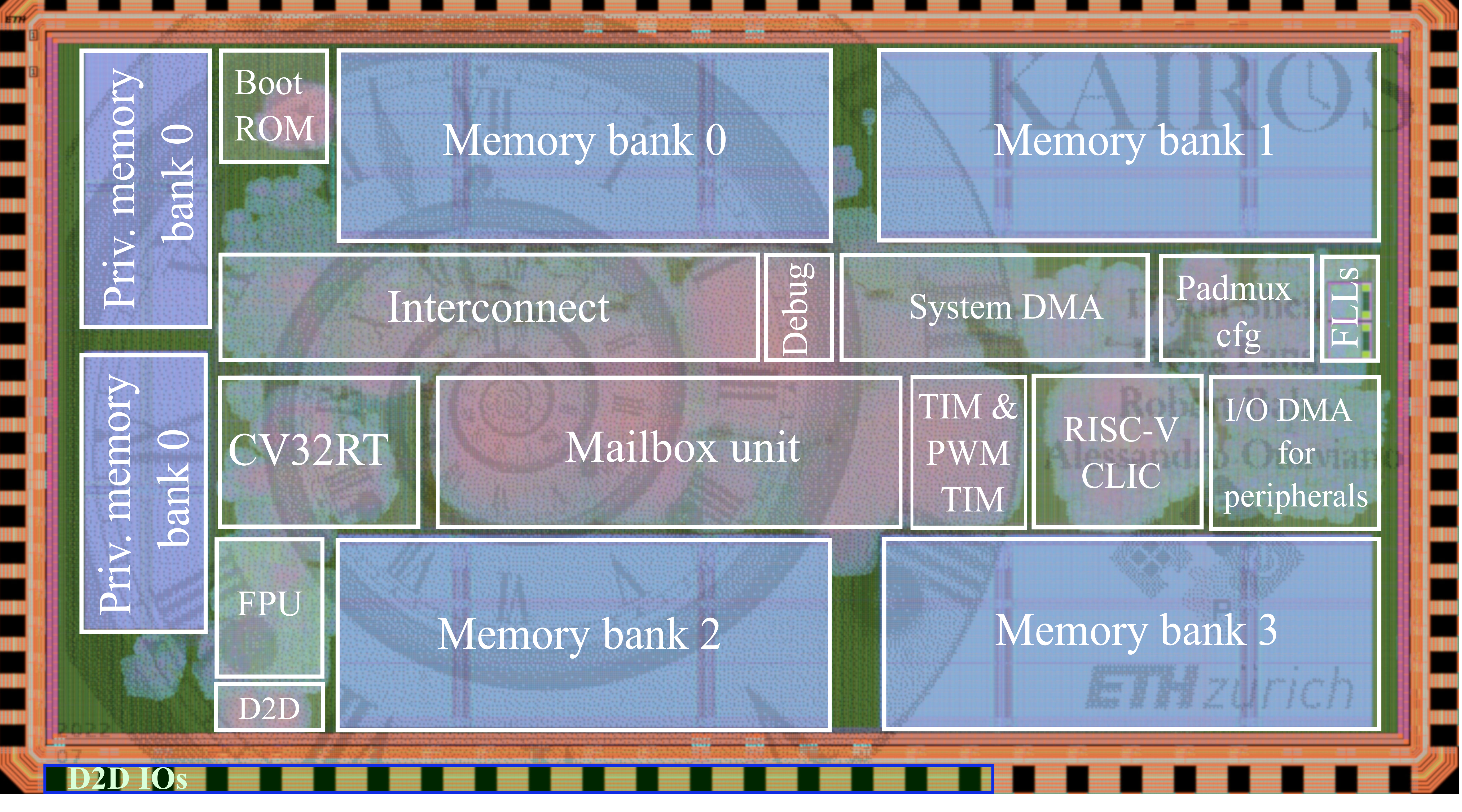}
            \label{subfig:kairos_micrograph}
        }
    \end{minipage}
    
    \vskip -0.4cm 

    \begin{minipage}[t]{\columnwidth}
        \renewcommand{\arraystretch}{0.93} 
        \resizebox{\linewidth}{!}{
        \begin{tabularx}{\linewidth}{X R}
            \arrayrulecolor{ieee-dark-black-100} \toprule
            \textbf{ISA} & rv32imafcxpulpv3xfastirq \\ 
            \textbf{Core} & CV32RT~\cite{balas2023cv32rt} \\
            \textbf{On-chip \gls{spm}} & \SI{448}{\kibi\byte} \\ 
            \arrayrulecolor{ieee-dark-black-40} \midrule
            \textbf{Technology} & {TSMC65} \\
            \textbf{Chip area} & \SI{7.2}{\square\milli\meter} \\
            \textbf{Logic size} & \SI{5}{\mega\GE} \\
            $\mathbf{V_{\mathbf{dd,core}}}$/$\mathbf{V_{\mathbf{dd,IO}}}$ \textbf{(nom.)} & \SI{1.2}{\volt}/\SI{2.5}{\volt} \\
            \textbf{Frequency range} & \SIrange{20}{380}{\mega\hertz} \\
            \textbf{Peak Power} & $<$ \SI{45}{\milli\watt} \\ 
            \arrayrulecolor{ieee-dark-black-100} \bottomrule
        \end{tabularx}
        }
        \label{subfig:table1}
    \end{minipage}
    
    \vskip 0.1cm 

    \caption{Annotated die shot of Kairos with key characteristics. \revrej{We highlight the \gls{d2d} logic and IOs on the bottom left of the die. The IO pads are on the sides of the chip since the package is \gls{qfn}.}}
    \label{fig:kairos-silicon-demonstrator}
\end{figure}

\begin{figure}[t]
    \centering
    \includegraphics[width=.9\columnwidth]{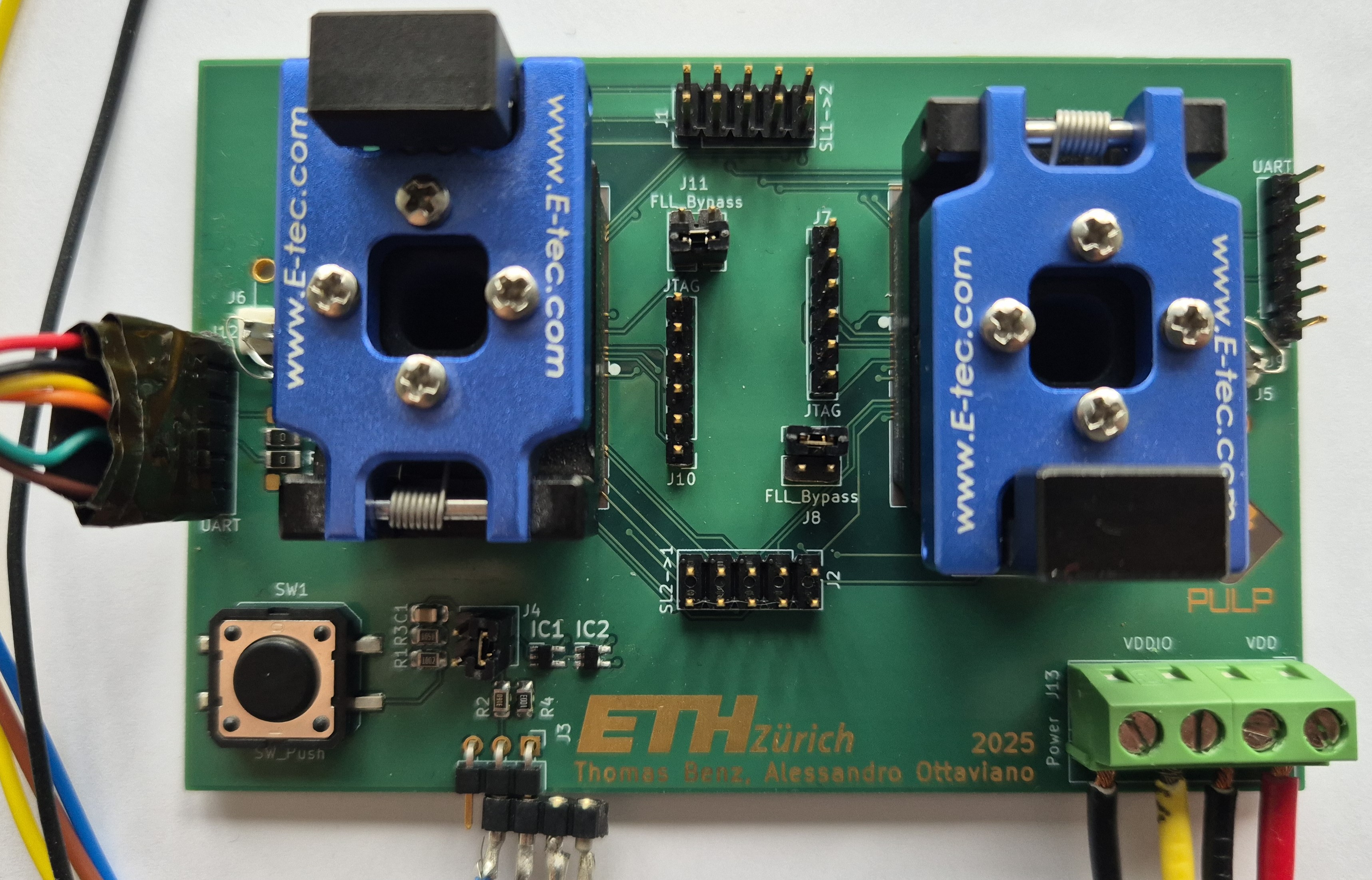}
    \caption{%
        \revrej{Kairos evaluation board. Two Kairos chip instances (each in one of the blue sockets) are connected to each other on the PCB through the \gls{d2d} link. \Gls{d2d} probes are visible on the top and bottom of the figure.}
    }
    \label{fig:eval:kairos-eval-board}
\end{figure}

We evaluate ControlPULPlet's area and energy footprint through a silicon demonstrator named \emph{Kairos}, which is designed, implemented, and fabricated using {TSMC}'s \SI{65}{\nano\meter} node, targeting a \SI{200}{\mega\hertz} system clock in the SS corner at \SI{125}{\celsius}.
\Cref{fig:kairos-silicon-demonstrator} shows Kairos's annotated die shot and implementation characteristics. 
The chip area is \SI{7.2}{\square\milli\metre}, housed in a \gls{qfn}64 package with nominal core and IO voltages $V_{\text{dd,core}}$=\SI{1.2}{\volt} and $V_{\text{dd,IO}}$=\SI{2.5}{\volt}, respectively.
\revrej{All of Kairos' external pads are implemented as regular IOs.}

We configure Kairos without \gls{dsa} ports due to area constraints, as the \gls{pmca} is not the primary focus of this work and has already been thoroughly verified in~\cite{Ottaviano2024}.
The manager domain features \SI{448}{\kibi\byte} of \gls{spm} to accommodate control firmware components and libraries. 
The system interconnect configuration defaults to \SI{32}{\bit} data and \SI{32}{\bit} addresses, while the \gls{dma} is configured with a data width of \SI{64}{\bit}.
The chip integrates one system and PWM timers, one UART, four I2C, four SPI hosts, and 64 mailboxes of \SI{32}{\byte} each.
We include two on-chip \glspl{fll} for the manager and peripheral domain clocks.
The \gls{d2d} link is configured with \tvlsirev{$CH=1$} and \tvlsirev{$LN=8$}, totaling $N_{wrs}=18$, and \tvlsirev{$CRD=8$}.

\paragraph*{\revrej{\textbf{Limitations}}}
\revrej{We are constrained to a low-performance configuration due to the 64-pin capacity of the surface-mounted, wire-bonded \gls{qfn}64 package, reflecting practical limitations in accessing advanced 2.5D integration technologies.
This limitation constrained the design of the standalone evaluation board for \gls{d2d} link assessment, which is shown in~\cref{fig:eval:kairos-eval-board}.
The board integrates two Kairos chips, interconnected to enable communication via their respective \gls{d2d} links.
One chip serves as the controller, the other as a subordinate memory endpoint simulating the controlled system domain.}
\revrej{The packaging limitations primarily affect $T_{\delta}$ (see \cref{fig:d2d-link-chronograms}) due to the inter-chip track length, on average \SI{70.5}{\milli\meter} long and \SI{0.2}{\milli\meter} wide in our design compared to the $\approx$~\SI{1}{\milli\meter} of a native interposer-based design~\cite{BOW}.
$T_{\text{mem}}$ depends instead on the endpoint of the receiving \gls{d2d} link, typically the controlled system, and is not affected by the packaging choice. 
In this implementation, both chips share the same \gls{spm}-based memory system with single-cycle access interconnect, bounding $T_{\text{mem}}$ to 1--2 clock cycles.
\tvlsirev{An additional limitation arises from the use of a mature technology node, which restricts the minimum operating voltage and keeps the energy-per-bit higher than what advanced nodes could achieve. This partially obscures the performance benefits of our design, particularly the near-lossless interface compared to the monolithic AXI4 counterpart.}
Despite these limitations, the main \gls{d2d} link metrics within the chip boundaries (e.g., power, interface energy, and area) remain valid under our simpler off-chip packaging. 
We detail this evaluation in \cref{subsec:eval:silicon_perf}.}

\subsubsection{\tvlsirev{D2D Link Scalability and Multi-chiplet Support}}
\tvlsirev{This work evaluates the D2D interface using a two-chiplet setup, sufficient to characterize the source-synchronous D2D module in latency, energy, and throughput. Nonetheless, broader scalability and multi-chiplet integration are also relevant, especially for \emph{control-centric} applications.}
\tvlsirev{Two integration models arise in such systems. In a centralized setup, a manager controller chiplet connects to multiple controlled chiplets with distributed sensors/actuators, forming a hub-and-spoke topology. The main controller instantiates multiple AXI4 manager ports and D2D links. ControlPULPlet’s RT-DMA and AXI pipelining allow issuing multiple outstanding transactions in parallel. Since this topology increases area overhead, it underscores the importance of a compact D2D bridge.}
\tvlsirev{In the distributed model, each chiplet combines control and sensing/actuation logic, coordinating via D2D links and shared memory (e.g., mailbox or local SPM). The number of D2D ports per chiplet defines the topology, from ring (two) to mesh (four). For control use cases, ring-based topologies are favored due to minimal controller interference.}

\subsubsection{\tvlsirev{Adaptability to Advanced Packaging Technologies}}

\tvlsirev{Our D2D interface is agnostic to the underlying packaging technology and requires only a routed signal path with controlled impedance and bounded clock skew, constraints that are met by several advanced packaging platforms. In particular, EMIB and Si-IF are well-matched to our D2D design: EMIB provides short silicon bridges with low trace length and skew, ideal for moderate-bandwidth, source-synchronous links; Si-IF offers ultra-dense, low-latency connections, making it especially suitable for tightly coupled chiplets. CoWoS, while offering higher bandwidth and full passive interposers, may exceed the form factor and cost requirements typical of embedded control systems. InFO technologies support compact, low-profile integration for consumer or IoT devices, and our D2D link can be conservatively configured (e.g., fewer channels or reduced frequency) to remain within its timing margins. Overall, the electrical assumptions of our link are compatible with the wiring models of these platforms, enabling straightforward integration across a range of packaging options.}

\begin{figure}[t]
    \centering
    \includegraphics[width=\columnwidth]{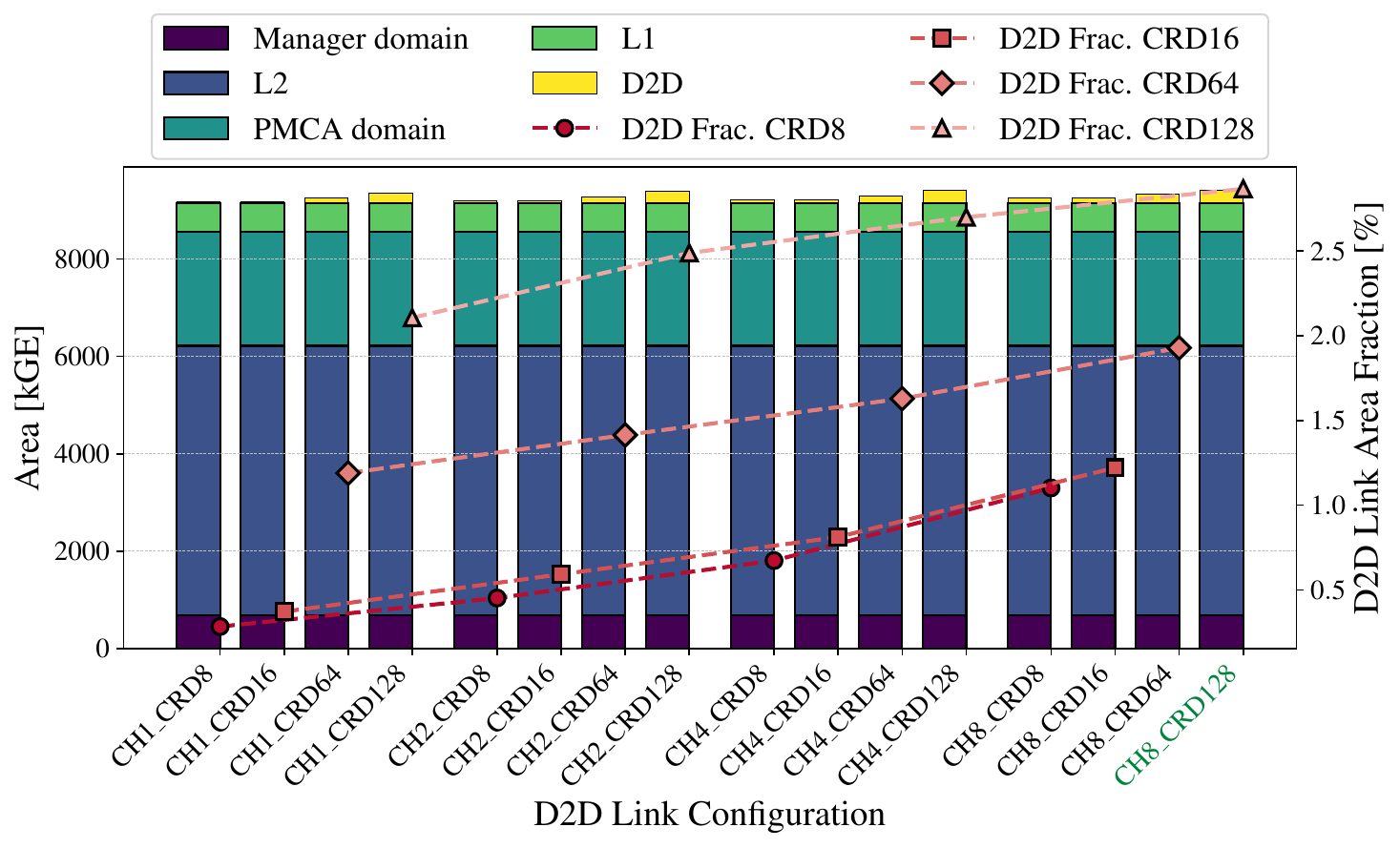}
    \caption{%
         Impact of \gls{d2d} link at varying channels and flow control FIFO depth on ControlPULPlet area. \texttt{LN} is fixed at \SI{8}{\bit}.
    }
    \label{fig:eval:cpulplet-sl-sweep}
\end{figure}

\begin{figure}[t]
    \centering
    \includegraphics[width=\columnwidth]{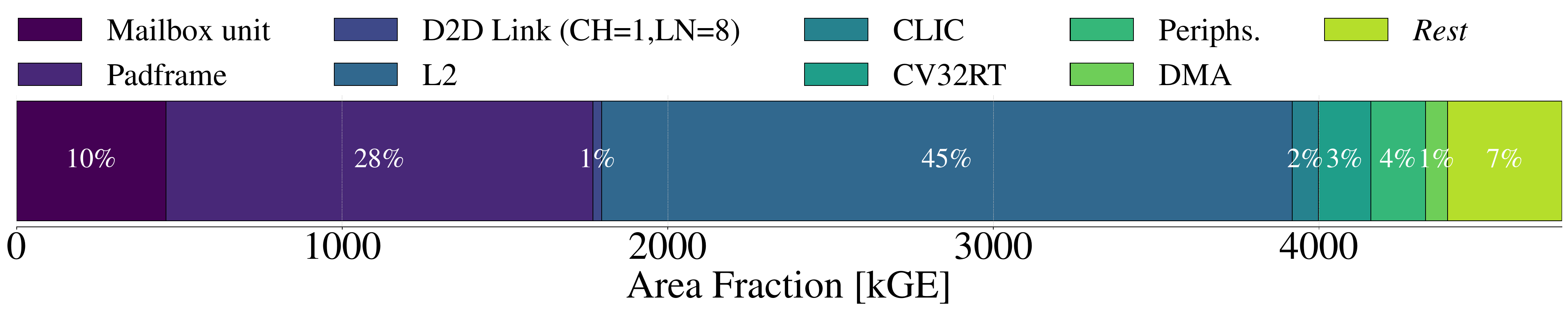}
    \caption{%
        Kairos area breakdown. \emph{Rest} denotes the interconnect system, its adapters and other logic.
    }
    \label{fig:eval:kairos-area-breakdown}
\end{figure}

\begin{figure}[t]
    \centering
    \includegraphics[width=\columnwidth]{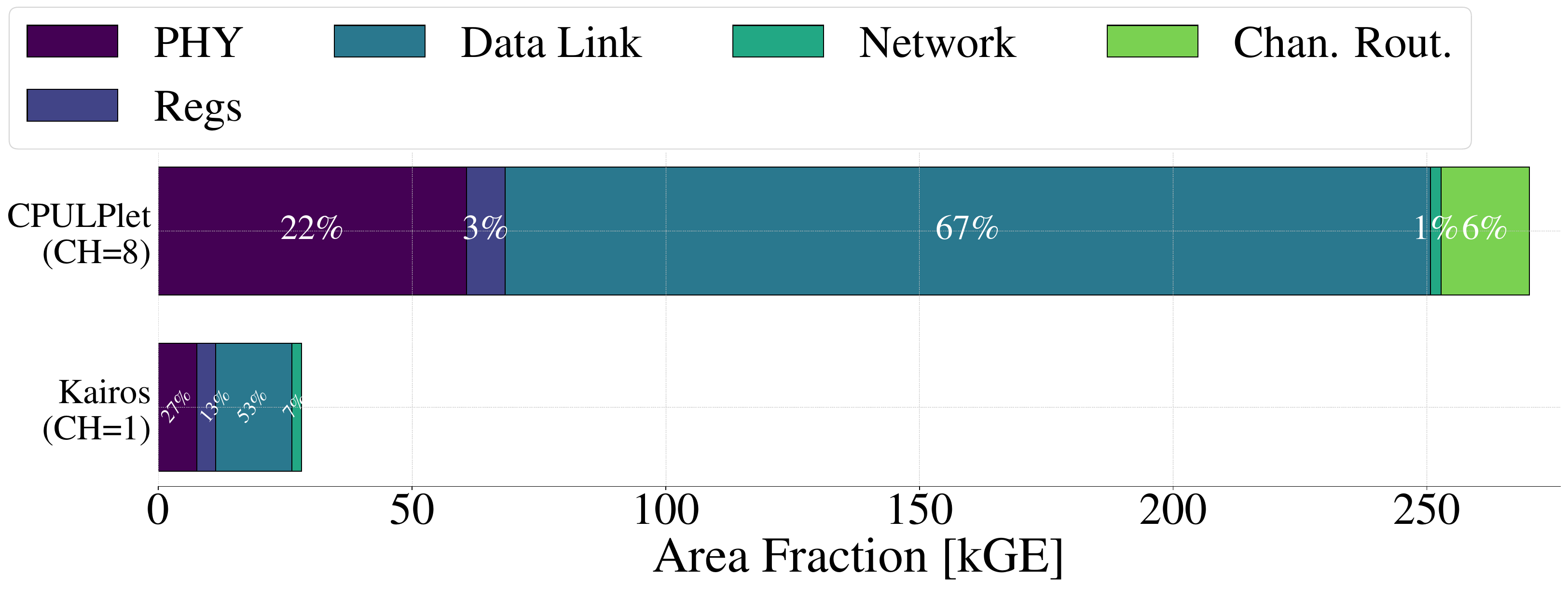}
    \caption{%
        \Gls{d2d} link area breakdown in ControlPULPlet (\tvlsirev{$CH=8$}, \tvlsirev{$CRD=128$}), and the Kairos chip demonstrator (\tvlsirev{$CH=1$}, \tvlsirev{$CRD=8$}).
    }
    \label{fig:eval:sl-area-breakdown}
\end{figure}

\begin{figure}[t]
    \centering
    \includegraphics[width=\columnwidth]{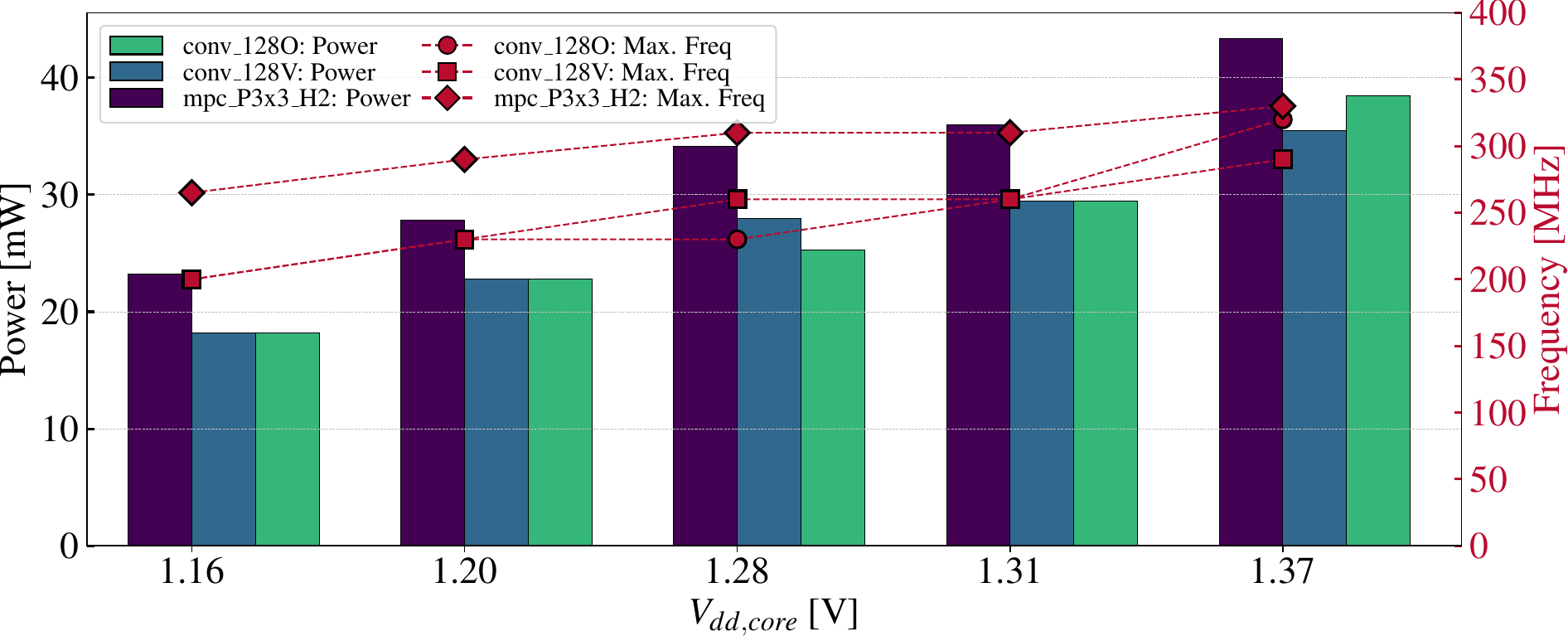}
    \caption{%
        Kairos power at maximum frequency and varying $V_{\text{dd,core}}$ for some relevant benchmarks.
    }
    \label{fig:eval:kairos-core-power}
\end{figure}

\begin{figure}[t]
    \centering
    \includegraphics[width=\columnwidth]{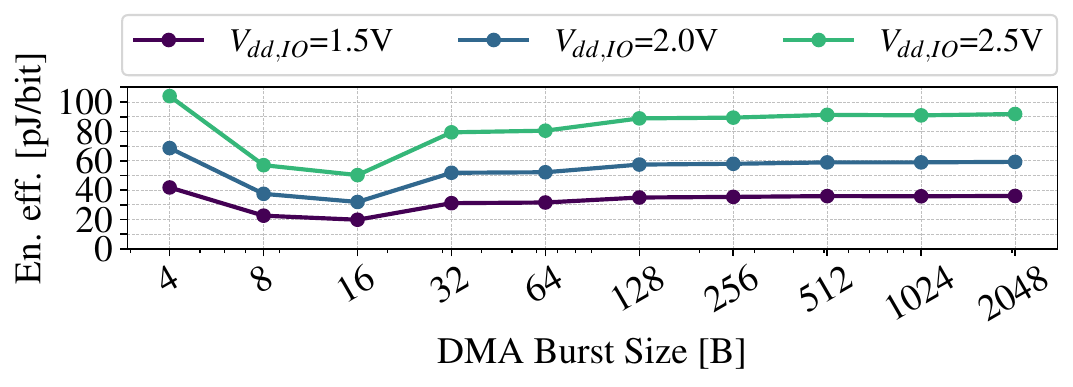}
    \caption{%
        \revrej{\Gls{d2d} energy efficiency in Kairos at different $V_{dd,IO}$ and \gls{dma} burst lengths, measured using the standalone board.}
    }
    \label{fig:eval:kairos-d2d-power}
\end{figure}

\subsection{Silicon Performance}\label{subsec:eval:silicon_perf}

We first explore the absolute and relative area impact of the \gls{d2d} in ControlPULPlet, including the manager and \gls{pmca} domains, under various parametrizations using post-synthesis data. 
Next, we examine the fabricated chip demonstrator configuration (manager domain only), detailing the overall area and energy figures of its key components.

\subsubsection{Area breakdown} 

\Cref{fig:eval:cpulplet-sl-sweep} illustrates the impact of progressively increasing the \gls{d2d} link's $CH$ (from 1 to 8) and $CRD$ (from 8 to 128) parameters on ControlPULPlet's area in \si{\GE}.
A \si{\GE} represents the area of a two-input, minimum-strength {NAND} gate.
The manager domain \glspl{spm} and \gls{pmca} dominate the area cost, accounting for nearly \SI{60}{\percent} of the system.
The link area overhead at fixed flow control credits --- dashed red lines referring to the right y-axis --- increases approximately linearly, which is expected due to a linear increase in the number of channels.
We observe that the most performant configuration in terms of attainable bus utilization (i.e., \tvlsirev{$CH=8$} and \tvlsirev{$CRD=128$},~\cref{subsec:eval:bw_lat}) is lightweight, incurring just \SI{2.9}{\percent} area overhead in-system.

\cref{fig:kairos-silicon-demonstrator} and \cref{fig:eval:kairos-area-breakdown} highlight the approximate location of Kairos main components and their relative size in the fabricated chip, respectively. The IO pads are on the sides of the chip due to the \gls{qfn} package.
The memory system still dominates the overall area, while the minimal configuration of the link (\tvlsirev{$CH=1$}, \tvlsirev{$LN=8$}) incurs \SI{1}{\percent} area overhead.
The RT mid-end integrated into both system and IO \gls{dma} engines adds a minimal area overhead of \SI{2}{\kGE} compared to a vanilla \gls{dma}, primarily due to the additional configuration registers and period counters.
Similarly, the fast interrupt handling features impact just \SI{10}{\percent} on the area of the\tvlsirevdel{vanilla} CV32RT core.

\Cref{fig:eval:sl-area-breakdown} presents a detailed area breakdown of the \gls{d2d} link components in ControlPULPlet (top) and the Kairos demonstrator (bottom). In the former, the control flow FIFO occupies more than \SI{60}{\percent} of the \gls{d2d} link area.
As the number of channels increases, the PHY's area contribution starts dominating the \gls{d2d} link.
The size of a single PHY is independent of the flow control FIFO and is particularly lightweight at only \SI{7.6}{\kGE} per channel.
\tvlsirevdel{As discussed in~Sec. II-C, since the Kairos demonstrator has $CH=1$, it does not include the channel router module.}

\subsubsection{Power Consumption}

\paragraph{\textbf{System Level}}

%
To assess Kairos' power consumption, we select two representative scenarios with varying computational and memory demands (\cref{fig:eval:kairos-core-power}). 
The \texttt{conv\_128} benchmark performs a \SI{32}{\bit} integer convolution on a \texttt{128$\times$128} matrix using a \texttt{5$\times$5} filter. We evaluate a vanilla version (\texttt{conv\_128V}) and an optimized one with SIMD, post-increment load/store, and loop unrolling (\texttt{conv\_128O}).
The \texttt{mpc\_P3$\times$3\_H2} is that described in~\cref{subsec:eval:case_study}.
\Cref{fig:eval:kairos-core-power} shows the power consumption (left axis, in black) of the chip at various core voltages $V_{\text{dd,core}}$ around the nominal $V_{\text{dd,core}}$=\SI{1.2}{\volt}. 
For each operating voltage, we find and report the maximum operating frequency (right axis in red).
At the nominal \SI{1.2}{V}, Kairos can achieve frequencies up to \SI{290}{\mega\hertz}, while maintaining a power envelope under \SI{30}{\milli\watt}. A cross-comparison with \gls{sota} is provided in the following~\cref{sec:rel}.
The D2D link accounts for only \SI{1.2}{\percent} of total power in silicon-calibrated post-layout simulations using memory-bound synthetic tests.

\paragraph{\textbf{\revrej{\Gls{d2d} Link}}}

\revrej{The energy consumption per transferred bit of the \gls{d2d} interface is affected by two components: one on-chip, due to the \gls{d2d} microarchitecture, and one off-chip, due to PCB trace length and IO voltage. With \gls{d2d} interface, we intend the whole network, data link, and PHY modules.}
\revrej{We evaluate the first component with the testbench framework shown in~\cref{fig:eval:cpulplet-sim-loopback}, which provides fine-grained internal energy estimates within the chip, isolating the \gls{d2d} link from the rest of the SoC through silicon-calibrated post-layout simulations of ControlPULPlet. 
The second component is assessed with the manufactured board, as this setup accounts for the impact of I/O voltage and PCB track length between the two chip instances. 
In both cases, we use \gls{dma}-operated transfers to maximize bus utilization, as in \cref{subsec:eval:bw_lat}.}
\revrej{Silicon-calibrated post-layout simulations of the chip yield a total internal energy-per-bit of \SI{1.3}{\pico\joule\per\bit} at nominal $V_{\text{dd,core}}$=\SI{1.2}{\volt}, including network, data link, and physical layers.
The PHY accounts for roughly half of this (\SI{0.62}{\pico\joule\per\bit}), with the remainder spent on AXI4-to-AXIS conversion and flow control logic.}
\revrej{\Cref{fig:eval:kairos-d2d-power} reports the energy-per-bit across three $V_{\text{dd,IO}}$ values, from the minimum functional \SI{1.5}{\volt} to the nominal \SI{2.5}{\volt}, at varying \gls{dma} burst sizes. 
The energy is minimized at \SI{16}{\byte}, matching the data transferred per cycle in \gls{ddr} mode (\tvlsirev{$CH=1$}, \tvlsirev{$LN=8$}). 
Transfers of \SI{4}{\byte} are inefficient, as they use only half of the \gls{dma} bus width (\SI{8}{\byte}). 
For burst sizes from \SI{32}{\byte} to \SI{2}{\kibi\byte}, corresponding to 256 beats and the maximum supported by AXI4, energy increases due to burst fragmentation and flow control FIFO limitations.}
\revrej{At $V_{\text{dd,core}}$=\SI{1.5}{\volt} and recalling the average reach of \SI{70.5}{\milli\meter}, the minimum normalized energy is \SI{18.23}{\pico\joule\per\bit}, highlighting the penalty introduced by I/O voltage and PCB traces compared to internal \gls{d2d} energy.
A 2.5D integration would mitigate this penalty, as interposer tracks are almost two orders of magnitude shorter and offer better electrical properties than standard FR4 PCB traces.}

\section{Related Work}\label{sec:rel} 

\tvlsirevdel{This section reviews relevant controller architectures and D2D interfaces.} \tvlsirev{In the following,} \cref{subsec:relwrk:ctrllers} discusses controllers \revrep{across different topologies}{with different} \tvlsirev{acceleration and real-time features}, with a focus on their \gls{d2d} interface, when \tvlsirevrep{information is}{available}. \Cref{subsec:relwrk:d2d} then examines \tvlsirev{standalone} \gls{d2d} interfaces \tvlsirevrep{independently}{proposed by prior works}.

\newcommand{\rot}[1]{\rotatebox[origin=c]{90}{#1}}
\newcommand{\tilt}[1]{\hspace{-1cm}\rotatebox[origin=c]{32}{#1}}
\newcommand{\noc}{\textcolor{ieee-dark-grey-40}{0}}
\newcommand{\dk}{\textcolor{ieee-bright-red-100}{?}}
\newcommand{\na}{\textcolor{ieee-dark-grey-40}{n.a.}}
\newcommand{\nad}{\textcolor{ieee-dark-grey-40}{-}}
\newcommand{\dl}[2]{\makecell[cc]{#1 \\ #2}}
\newcommand{\dll}[2]{\makecell[cl]{#1 \\ #2}}
\newcommand{\dllb}[2]{\makecell[cl]{\textbf{#1} \\ \textbf{#2}}}
\newcommand{\dlb}[2]{\makecell[cc]{\textbf{#1} \\ \textbf{#2}}}
\newcommand{\tl}[3]{\makecell[cc]{#1 \\ #2 \\ #3}}
\newcommand{\tll}[3]{\makecell[cl]{#1 \\ #2 \\ #3}}
\newcommand{\tlb}[3]{\makecell[cc]{\textbf{#1} \\ \textbf{#2} \\ \textbf{#3}}}

\begin{table*}[ht]
    \centering
    \caption{Comparison of programmable controller classes. We highlight several metrics of the controller\tvlsirevdel{as a system and its D2D interface}. \tvlsirev{The autonomous DMA is not highlighted, as it is unique to our design.}}
    \label{tab:controller-comparison}
    \setlength{\tabcolsep}{1.5pt} 
    \renewcommand{\arraystretch}{1.1} 
    \resizebox{\linewidth}{!}{
    \begin{threeparttable}
    
    \begin{tabular}{lcccccccccccccc}
        \arrayrulecolor{ieee-dark-black-100} \toprule
        
        \multicolumn{9}{c}{} & \multicolumn{6}{c}{\textbf{D2D}} \\
        \arrayrulecolor{ieee-dark-black-40} \cmidrule(lr){10-15}
        
        \textbf{Controller} & 
        \textbf{Class} & 
        \textbf{Applic.} &
        \dlb{\Gls{dsa}}{Integr.} & 
        \dlb{\Gls{d2d}}{Interf.} & 
        \dlb{Fast}{Interr.} & 
        \dlb{Tech}{[\si{\nano\metre}]} &
        \dlb{Max. Freq.}{[\si{\mega\hertz}]} &
        \dlb{Peak Power}{[\si{\milli\watt}]} & 
        \textbf{Prot.} & 
        \textbf{\#Links} & 
        \textbf{\#Wires/Link} & 
        \dlb{Peak BW}{[\si{\giga\bit\per\second}]} &
        \dlb{Reach}{[\si{\milli\meter}]} &
        \dlb{En. eff.}{[\si{\pico\joule\per\bit}]} \\
        
        \arrayrulecolor{ieee-dark-black-40} \midrule

        \multicolumn{15}{c}{\textbf{\tvlsirev{Controllers for 2D integration with simple on-chip or off-chip interfaces}}} \\

        \arrayrulecolor{ieee-dark-black-40} \cdashline{1-15}

        \textbf{STM32}~\cite{STM32} & 
        1-core & 
        IoT &
        \textcolor{ieee-bright-dgreen-100}{\cmark} & 
        \textcolor{ieee-bright-red-100}{\xmark} & 
        \textcolor{ieee-bright-red-100}{\xmark} & 
        28 & 
        24-480 & 
        66-132 @\SI{3.3}{\volt} & 
        \nad & 
        \nad & 
        \nad &
        \nad &
        \nad &
        \nad \\

        \textbf{TC21XL}~\cite{AURIX_TC21XL} & 
        1-core & 
        automotive &
        \textcolor{ieee-bright-dgreen-100}{\cmark} & 
        \textcolor{ieee-bright-red-100}{\xmark} & 
        \textcolor{ieee-bright-red-100}{\xmark} & 
        65 & 
        133 & 
        290 @\SI{3.3}{\volt} & 
        \nad & 
        \nad & 
        \nad &
        \nad &
        \nad &
        \nad \\

        \textbf{PCU}~\cite{schone_energy_2019} & 
        \dl{1-core,}{FSMs} & 
        \dl{CPU}{DVFS} &
        \textcolor{ieee-bright-red-100}{\xmark} & 
        \textcolor{ieee-bright-red-100}{\xmark} & 
        \na & 
        \tnote{a}~\nad & 
        \na & 
        \na & 
        \nad & 
        \nad & 
        \nad &
        \nad &
        \nad &
        \nad \\

        \textbf{OCC}~\cite{ibm_occ} & 
        1-core & 
        \dl{CPU}{DVFS} &
        \textcolor{ieee-bright-red-100}{\xmark} & 
        \textcolor{ieee-bright-red-100}{\xmark} & 
        \textcolor{ieee-bright-red-100}{\xmark} & 
        \tnote{a}~\nad & 
        $<$266-658 & 
        \na & 
        \nad & 
        \nad & 
        \nad &
        \nad &
        \nad &
        \nad \\

        \textbf{SCP}~\cite{ARM_PCSA} & 
        1-core & 
        \dl{CPU}{DVFS} &
        \textcolor{ieee-bright-red-100}{\xmark} & 
        \textcolor{ieee-bright-red-100}{\xmark} & 
        \textcolor{ieee-bright-dgreen-100}{\cmark} & 
        \tnote{a}~\nad & 
        $<$400 & 
        \na & 
        \nad & 
        \nad & 
        \nad &
        \nad &
        \nad &
        \nad \\

        \textbf{TC29x}~\cite{AURIX_TC29X} & 
        3-cores & 
        automotive &
        \textcolor{ieee-bright-dgreen-100}{\cmark} & 
        \textcolor{ieee-bright-red-100}{\xmark} & 
        \textcolor{ieee-bright-dgreen-100}{\cmark} & 
        \nad & 
        $<$400 & 
        1697 @\SI{3.3}{\volt} & 
        \nad & 
        \nad & 
        \nad &
        \nad &
        \nad &
        \nad \\

        \textbf{AST2600}~\cite{ASPEED_AST2600} & 
        2-cores & 
        \dl{server}{mngmnt} &
        \textcolor{ieee-bright-red-100}{\xmark} & 
        \textcolor{ieee-bright-red-100}{\xmark} & 
        \textcolor{ieee-bright-red-100}{\xmark} & 
        28 & 
        \dl{1000, Cort. A}{200, Cort. M} & 
        $> 2000$ @1.8-3.3~\si{\volt} & 
        \nad & 
        \nad & 
        \nad &
        \nad &
        \nad &
        \nad \\

        \textbf{TC4x}~\cite{AURIX_TC4X} & 
        \dl{6-cores}{\glspl{dsa}} & 
        automotive &
        \textcolor{ieee-bright-dgreen-100}{\cmark} & 
        \textcolor{ieee-bright-red-100}{\xmark} & 
        \textcolor{ieee-bright-dgreen-100}{\cmark} & 
        28 & 
        500 & 
        2240-2501 @\SI{3.3}{\volt} & 
        \nad & 
        \nad & 
        \nad &
        \nad &
        \nad &
        \nad \\

        \textbf{ControlPULP}~\cite{Ottaviano2024} & 
        \dl{1-core}{\glsentrytext{pmca}} & 
        automotive &
        \textcolor{ieee-bright-dgreen-100}{\cmark} & 
        \textcolor{ieee-bright-red-100}{\xmark} & 
        \textcolor{ieee-bright-dgreen-100}{\cmark} & 
        \tnote{d}~22 & 
        \tnote{d}~500 & 
        \na & 
        \nad & 
        \nad & 
        \nad &
        \nad &
        \nad &
        \nad \\

        \arrayrulecolor{ieee-dark-black-40} \cdashline{1-15}

        \multicolumn{15}{c}{\textbf{\tvlsirev{Controllers for 2.5D integration with die-to-die interfaces}}} \\

        \arrayrulecolor{ieee-dark-black-40} \cdashline{1-15}

        \textbf{AMD SMU}~\cite{AMD_SMU,AMD_ROME} & 
        1-core & 
        \dl{CPU}{DVFS} &
        \textcolor{ieee-bright-red-100}{\xmark} & 
        \textcolor{ieee-bright-dgreen-100}{\cmark} & 
        \textcolor{ieee-bright-red-100}{\xmark} & 
        \tnote{b}~12 & 
        \na & 
        \na & 
        IFOP & 
        \tnote{c}~8 & 
        \na &
        \dl{\tnote{c}~3520}{@\SI{2}{\giga\hertz}} &
        \na &
        \dl{\tnote{c}~2}{@\SI{1.2}{\volt}} \\

        \textbf{Liu~\etal}~\cite{CICC} & 
        1-core & 
        \na &
        \textcolor{ieee-bright-red-100}{\xmark} & 
        \textcolor{ieee-bright-dgreen-100}{\cmark} & 
        \textcolor{ieee-bright-red-100}{\xmark} & 
        16 & 
        \dl{500, OpenRISC}{1000, \gls{d2d}} & 
        \na & 
        AIB-like & 
        3 & 
        96 &
        \dl{2 (per pin)}{@\SI{1}{\giga\hertz}} &
        $<$2 &
        \dl{1.48}{@\SI{1.2}{\volt}} \\

        \textbf{Vivet~\etal}~\cite{INTACT} & 
        \na & 
        \dl{CPU}{DVFS} &
        \na & 
        \textcolor{ieee-bright-dgreen-100}{\cmark} & 
        \na & 
        \na & 
        1200 & 
        \tnote{e}~1160 & 
        3D Plug & 
        6+8+1 & 
        12 &
        \dl{3152}{@\SI{1.1}{\giga\hertz}} &
        1.5-1.8 &
        \dl{0.75}{@\SI{1.2}{\volt}} \\

        \dll{\textbf{ControlPULPlet}}{\textbf{(ours)}} & 
        \textbf{\dl{1-core}{\gls{pmca}}} & 
        \na &
        \textcolor{ieee-bright-dgreen-100}{\cmark} & 
        \textcolor{ieee-bright-dgreen-100}{\cmark} & 
        \textcolor{ieee-bright-dgreen-100}{\cmark} & 
        \textbf{65} &  
        \tnote{f}~\textbf{290} & 
        \textbf{\tnote{f}~30 / \tnote{g}~210} & 
        \textbf{custom} & 
        \textbf{\tnote{f}~1 -- \tnote{h}~8} & 
        \textbf{16+2} &
        \textbf{\dl{\tnote{h}~$>$51}{@\SI{200}{\mega\hertz}}} &
        \textbf{$<$70.5} &
        \dlb{\tnote{i}~1.3 -- \tnote{f}~11.7}{\tnote{l}~@\SI{1.2}{\volt}} \\

        \arrayrulecolor{ieee-dark-black-100} \bottomrule
    \end{tabular}

    \begin{tablenotes}[para, flushleft]
        \fontsize{7.7pt}{7.7pt}\selectfont
        \item[a] soft IP not tied to standalone chip
        \item[b] IO die (IOD)
        \item[c] from AMD Rome; we report Infinity Fabric numbers, exact \glsentrytext{scf} values are unavailable
        \item[d] soft IP not tied to standalone chip, reported post-synthesis estimates
        \item[e] one SCVR
        \item[f] Kairos demonstrator @$V_{\text{dd,core}}$=\SI{1.2}{\volt}, $V_{\text{dd,IO}}$=\SI{1.5}{\volt}
        \item[g] considering both manager and \gls{pmca} domains
        \item[h] peak with \texttt{CH}=8, \texttt{CRD}=128, \SI{90}{\percent} of the on-chip control interface requirements; optimizable at additional area cost
        \item[i] internal energy estimated with silicon-calibrated post-layout simulation
        \item[l] scaled from $V_{\text{dd,IO}}$=\SI{1.5}{\volt} (measured on Kairos demonstrator)
    \end{tablenotes}
    \end{threeparttable}
    }
\end{table*}

\subsection{\revrej{Programmable Hardware Controllers}}\label{subsec:relwrk:ctrllers}

\tvlsirevrep{Existing controllers can be classified into two categories: hardwired and programmable. Hardwired controllers rely on state machines and timers for fixed-function control. They offer the highest speed and lowest silicon cost, but lack flexibility, a limitation addressed by programmable controllers.}{We focus on \textit{programmable} controllers, which address the limited flexibility of fixed-function controllers~\cite{CONTROLLER_UMBRELLA_TCASII_1,GANSI_1}}.
\tvlsirevrep{The latter}{Programmable controllers} are classified as (i) \revrep{single-core}{\emph{single-core}} programmable, (ii) \revrep{multi-core}{\emph{multi-core}} programmable, and (iii) \revrep{heterogeneous multi-core}{\emph{heterogeneous multi-core}} controllers with \glspl{dsa}.
\Cref{tab:controller-comparison} highlights the relevant \gls{sota} in this category\tvlsirev{. The table groups controllers by D2D interface support.}

\paragraph{\textbf{\revrej{Single-core Programmable Controllers}}}

\revrej{Single-core programmable controllers like STMicroelectronics' \SI{32}{\bit} STM32 series (\cref{fig:ctrl-types}a) support real-time applications with accelerated context switching, consuming less than \SI{150}{\milli\watt} at \SI{500}{\mega\hertz}~\cite{STM32}. 
Similarly, AURIX TriCore TC2xx series provides hardware-assisted context switching\rev{with 10-16 clock cycles interrupt response time and 156-162 clock cycles context switching}~\cite{AURIX_TC21XL}, \revdel{but focuses less on low power consumption,}consuming \SI{300}{\milli\watt} at just \SI{130}{\mega\hertz}.}\tvlsirev{Our design outperforms TriCore in interrupt latency and context switching (both 1.6$\times$ faster) with similar power consumption.}

\revrej{On-chip single-core controllers for CPU \gls{dvfs} \rev{and SoC management} include Intel’s power control unit (PCU)~\cite{schone_energy_2019}, IBM’s on-chip controller (OCC)~\cite{ibm_occ}, Arm’s system control processor (SCP)~\cite{ARM_PCSA}, and AMD’s system management unit (SMU)~\cite{AMD_SMU}, all based on \SI{32}{\bit} commercial processors.
These controllers implement simple PID control policies with control periods of \SIrange{0.25}{1}{\milli\second}, aligning favorably with ControlPULP's capabilities\tvlsirevdel{ explored in [3]}. \tvlsirevdel{However, no data is provided on the execution time of advanced predictive policies, which remain uncommon in embedded controllers used in industry and are typically limited to software implementations running on the PEs themselves.}
The SMU uniquely integrates a chiplet-based \gls{d2d} interface via AMD's Infinity Fabric on Package (IFOP). 
IFOP \revrep{integrates}{comprises} a scalable data and control fabric (SDF and SCF, respectively). Online CPU \gls{dvfs}\revdel{control} {is} managed through the SC\tvlsirevrep{F}{P} with a \SI{1}{\milli\second}-period PID policy, facilitating communication from the SMU, located within the IO die (IOD) chiplet, to the compute chiplets housing high-performance Zen cores. 
In AMD's Rome~\cite{AMD_ROME}, IFOP achieves \SI{2}{\pico\joule\per\bit} and \SI{440}{\giga\byte\per\second} at \SI{1.2}{\volt} on a \SI{12}{\nano\metre} passive interposer, though control and data planes likely have different bandwidth needs. 
A similar chiplet-compatible \gls{d2d} demonstrator using an OpenRISC microcontroller in \SI{16}{\nano\metre} achieves \SI{2}{\giga\bit\per\second} per pin with \SI{1.5}{\pico\joule\per\bit} at \SI{1.2}{\volt}~\cite{CICC}. 
\revrep{These}{While these} controllers prioritize deterministic execution\rev{, fast context switching,} and \revrep{reliable}{efficient} \gls{d2d} \revrep{links rather than}{control interfaces, they lack the} computational power \revrep{for}{required by} advanced control algorithms.}

\paragraph{\textbf{\revrej{Multi-core Controllers}}}
\revrej{Multi-core controllers address this computational gap. The six-core AURIX TC29x~\cite{AURIX_TC29X} and ASPEED's AST2600, a common Baseboard Management Controller~\cite{ASPEED_AST2600}, demonstrate increased processing capability, reflected in their high power consumption\rev{,} between \SIrange{1.5}{2}{\watt}.
However, \revdel{their reliance on} general-purpose cores \revrep{results}{result} in high silicon cost and lower energy efficiency for specialized workloads.}

\paragraph{\textbf{\revrej{Heterogeneous Multi-core Controllers}}}
\revrej{To improve efficiency, heterogeneous multi-core controllers integrate \glspl{dsa}. 
ControlPULP combines a \SI{32}{\bit} embedded core with an 8-core tightly-coupled \gls{pmca}, balancing flexibility, performance, and efficiency for real-time control applications~\cite{Ottaviano2024}. \tvlsirevdel{It features an AXI4 interface for on-chip control and fast interrupt handling via a RISC-V CLIC.}
Similarly, AURIX TC4x series integrates a 6-core high-end controller with \rev{diverse} accelerators \rev{for AI-based model predictive and model adaptive control\tvlsirevdel{ (MPC and MAC, respectively), which ease vehicle trajectory control/planning, motor control, and radar pre-processing}}~\cite{AURIX_TC4X}.
\revrep{and r}{R}egister banking \revrep{for}{enables} fast interrupt response \rev{clock cycles and context switching, key properties retained from its predecessors}. \revrep{and, though at a higher power envelope $<$2.5W}{Increased acceleration and performance incurs additional power consumption costs than the low-end TC2x series, about 2.5W.}}
\revrej{Most multi-core and heterogeneous controllers are available as standalone chips or soft \gls{ip} blocks for on-chip integration, but few support chiplet-compatible \gls{d2d} interfaces. 
A notable exception is the 96-core demonstrator by Vivet~\etal, which integrates a power management controller \rev{including switched-capacitor voltage regulators (SCVRs), one for each compute chiplet,} on an active silicon interposer~\cite{INTACT}. 
\tvlsirevdel{The power manager hub for one compute chiplet occupies 11.4mm2 in 65nm CMOS, while its custom 3D Plug D2D interface achieves 394GB/s at 1.1GHz, with 0.75pj/b at 1.2V.}}

\begin{table*}[ht]
    \centering
    \caption{Comparison of standalone D2D Interfaces. We highlight several metrics of each interface. \emph{A} and \emph{D} refer to analog and digital, respectively.}
    \label{tab:d2d-comparison}
    \setlength{\tabcolsep}{1.5pt}
    \renewcommand{\arraystretch}{1.1}
    \resizebox{\linewidth}{!}{
    \begin{threeparttable}
    
    \begin{tabular}{lcccccccc}
    
        \arrayrulecolor{ieee-dark-black-100} \toprule
                
        \textbf{D2D Interface} & 
        \dlb{Tech}{[\si{\nano\metre}]} & 
        \dlb{On-chip}{Protocol} &
        \textbf{Design} &
        \dlb{Reach}{[\si{\milli\meter}]} &
        \dlb{Peak BW}{[\si{\giga\bit\per\second}]} &
        \dlb{Area}{[\tvlsirev{kGE}]} &
        \dlb{Energy}{[\si{\pico\joule\per\bit}]} &
        \dlb{\#Wires}{/Link} \\
        
        \arrayrulecolor{ieee-dark-black-40} \midrule
        

        \tvlsirev{\textbf{Kuttappa~\etal}~\cite{CICC25-3-INTEL}} & \tvlsirev{16} & \tvlsirev{\tnote{a}~AXI~(\SI{36}{\bit})} & \tvlsirev{A/D} & \tvlsirev{1.2} & \tvlsirev{\tnote{b}~18 @\SI{1}{\giga\hertz}} & \tvlsirev{\na} & \tvlsirev{\tnote{c}~0.85} & \tvlsirev{18+8} \\

        \tvlsirev{\textbf{Melek~\etal}~(UCIe)~\cite{ISSCC25-2}} & \tvlsirev{3} & \tvlsirev{\na} & \tvlsirev{A/D} & \tvlsirev{1.4} & \tvlsirev{2048 @8~\si{\giga\hertz}} & \tvlsirev{\na} & \tvlsirev{0.29 @0.45~\si{\volt}} & \tvlsirev{64+4} \\
        
        \textbf{BoW}~\cite{BOW} & 65-5 & \dl{PCIe, MAC}{PIPE/LPIF} & D & 10 & \tnote{d}~5-8 @2.5-4~\si{\giga\hertz} & \na & $<$1 @0.7-0.9~\si{\volt} & 16+2 \\
        
        \textbf{Lin~\etal}~\tvlsirev{(LIPINCON)}~\cite{LIPINCON} & 7 & Mesh (\SI{1968}{\bit}) & A/D & 0.5 & 2560 @4GHz & \tnote{e}~18420 & 0.56 @0.3~\si{\volt} & \tvlsirevrep{\na}{40+6} \\
        
        \textbf{GRS}~\cite{GRS} & 16 & \na & A/D & 80 & \tnote{d}~25 @12.5GHz & \tvlsirev{589.9} & 1.17 @0.3~\si{\volt} & 8+1 \\
        
        \textbf{Li~\etal}~\cite{CHIPLET_UMBRELLA} & Kintek-7 FPGA & \na~(\SI{64}{\bit}) & D & \na & 8.3 @\SI{250}{\mega\hertz} & \na & \na & 32+1+1 \\
        
        \textbf{\emph{Ours}} & \textbf{65} & \textbf{AXI4 (64b)} & \textbf{D} & \tnote{f}~\textbf{70.5} & \textbf{51 @\SI{200}{\mega\hertz}} & \textbf{\dl{\tnote{g}~\tvlsirev{28.2}/\tnote{h}~\tvlsirev{270.2}}{\tvlsirev{(7.6 one PHY)}}} & \textbf{\dl{11.7 \tvlsirev{(1.3 int.)}}{@\SI{1.2}{\volt}}} & \textbf{16+2} \\

        \arrayrulecolor{ieee-dark-black-100} \bottomrule
    \end{tabular}

    \begin{tablenotes}[para, flushleft]
        \fontsize{7.7pt}{7.7pt}\selectfont
        \item[a] The exact AXI protocol (AXI3 or AXI4) not specified
        \item[b] Aggregate BW of one cluster only
        \item[c] Operational voltage is not specified
        \item[d] Per pin
        \item[e] \tvlsirev{one channel (implemented 2 channels per chiplet)}
        \item[f] Reach based on the average trace length of the PCB shown in~\cref{fig:eval:kairos-eval-board}; other works refer to interposer-based reach
        \item[g] Network, data, and PHY \tvlsirev{for the Kairos config. $CH=1$, $LN=8$}
        \item[h] Network, data, and PHY \tvlsirev{for the peak BW config. $CH=8$, $LN=8$}
    \end{tablenotes}
    \end{threeparttable}
    }
\end{table*}

\subsection{\revrej{\Gls{d2d} Interfaces}}\label{subsec:relwrk:d2d}

\Cref{tab:d2d-comparison} compares industrial and academic \gls{d2d} protocols. 
\tvlsirev{While energy-per-bit is difficult to compare due to differing technology nodes and voltages, peak bandwidth can be scaled assuming higher frequencies, though still bounded by the frontend bus width. Area is reported in gate equivalents, normalized to a minimum-strength NAND gate, making it technology-agnostic.}
UCIe~\cite{ISSCC25-2}, BoW~\cite{BOW}, and LIPINCON \tvlsirev{over CoWoS}~\cite{LIPINCON} support \rev{widely used frontend standards, such as} PCIe and CXL. 
The standalone D2D interfaces and protocols surveyed in Table II \tvlsirev{are specifically optimized for high-throughput communication across the data plane of the chiplet, driving the design toward high-frequency operation (in the multi-\si{\giga\hertz} range) and the scaling out of individual link instances. \Cref{tab:d2d-comparison} presents the number of wires per link as a normalized figure of merit for this purpose.}
For example, GRS\cite{GRS} achieves \SI{25}{\giga\bit\per\second} per pin over an \SI{80}{\milli\metre} reach on a silicon interposer, operating at a high frequency of \SI{12.5}{\giga\hertz}.
\tvlsirevrep{This design choice is confirmed by the large data width of certain implementations like LIPINCON (almost 2000 parallel bits), and contrasts with the 64b employed in this work for the design of the control plane interface.}{Similarly, LIPINCON~\cite{LIPINCON} integrates 1968 parallel bits, while in~\cite{ISSCC25-2}, Cadence proposes a UCIe-compatible \gls{d2d} interface with 64 parallel data wires across 8 links delivering \SI{2048}{\giga\bit\per\second} at \SI{8}{\giga\hertz}.}
\tvlsirev{These implementations contrast with the \SI{64}{\bit} AXI4 frontend used in this work for the control plane interface, where high bandwidth and speed are not primary requirements. However, because of its flexibility, our \gls{d2d} design can be readily scaled to meet the demands of a wider data bus for data-intensive applications.}
\tvlsirev{Similar to our work, Intel’s design in~\cite{CICC25-3-INTEL} features a D2D interface with an AXI frontend, and a \SI{36}{\bit} data bus. The mixed analog/digital PHY is integrated into a 14-cluster \gls{soc} for media and ML applications, using 9 data wires per cluster to deliver \SI{18}{\giga\bit\per\second} at \SI{1}{\giga\hertz} per cluster.}
\tvlsirev{Though currently dataplane-focused, these high-bandwidth links enable future D2D communication in complex distributed control systems, making them increasingly relevant for control-plane use.}
\tvlsirevdel{To the best of the authors knowledge, none of the surveyed D2D interfaces was implemented or evaluated specifically targeting control applications.}
In academia, Li~\etal~\cite{CHIPLET_UMBRELLA} propose a custom\rev{, parallel} \gls{d2d} interface achieving \SI{8.3}{\giga\bit\per\second} at \SI{250}{\mega\hertz}. 
\rev{The microarchitecture only supports a fixed 32 parallel data lanes with a 64b buffer FIFO used for protocol conversion. Moreover, the D2D physical implementation assessment is limited to FPGA mapping without silicon prototyping, which prevents accurate power and area estimates.} 
\revdel{Their FPGA-based approach limits power, energy, and area characterization.}

\ifx\showtvlsirebuttal\undefined
\else
\subsection{\sout{Comparison and Discussion}}
\paragraph{\textbf{\sout{Controller-level}}}
\fi
\tvlsirevdel{Of the controllers shown in Table I, ControlPULPlet shares most design similarities with the AURIX series.} 
\tvlsirevdel{While our design is faster in terms of interrupt response time and context switching thanks to its hardware optimizations, achieving as low as 6 and 100 clock cycles compared to 10-16 and 156-162 of the TriCore processors, it is also competitive in terms of power envelope, achieving flexible PMCA-enabled acceleration at $<$250mW, almost 8$\times$ and 11$\times$ lower than the TC29x and TX4x, respectively. Moreover, although the AURIX processors integrate DMA engines with reliability features and independent control of transfer queues, no explicit information is reported on hardware-assisted periodic DMA transfers to fully eliminate processor context switch during periodic execution.}

\ifx\showtvlsirebuttal\undefined
\else
\paragraph{\textbf{\sout{D2D Link}}}
\fi
\tvlsirevdel{Regarding the D2D interface, o}Our \tvlsirev{D2D} design stands out for its lightweight area overhead\tvlsirev{, even when configured for peak throughput ($CH=8$, $LN=8$ for a data bus of 64 b)}. As shown in Table II, it achieves a smaller die size than PHY-only reported implementations in more advanced technology nodes, such as \cite{GRS} (2.2$\times$) and \cite{LIPINCON} (136$\times$), while also incorporating the data link and network layers.
\tvlsirevrep{On the throughput side}{Furthermore}, \tvlsirevdel{the high flexibility of our design allows to meet the requirements of the native AXI4 on-chip control interface. Furthermore, }the link achieves higher throughput compared to designs with similar microarchitecture and operating frequency, e.g.,~\cite{CHIPLET_UMBRELLA}, which reports 6$\times$ less bandwidth while doubling the number of wires per link.
\tvlsirev{At iso-frequency, our interface achieves superior peak bandwidth, e.g., outperforming the recent~\cite{CICC25-3-INTEL} by $14 \times$, thanks to the high bus utilization enabled by the D2D design, despite other works using more advanced technology nodes.}
The energy efficiency of our silicon prototype is negatively affected by the flat QFN package and the lack of an interposer as substrate, leading to PCB-induced overhead (PCB track parasitic).
However, cycle-accurate simulations provide high observability of the inner D2D circuitry, enabling silicon-calibrated measurement of its internal energy (1.3~pJ/b), a lower bound independent from the length and parasitics of the off-die wires. The inner D2D energy is comparable with leading academic and industry designs at the same voltage, e.g.,~\cite{CICC} (1.48 pJ/b) and the AMD SMU~\cite{AMD_SMU} (2 pJ/b).
\tvlsirev{Moreover, since energy-per-bit scales with the square of the operating voltage, a more advanced technology node enabling lower voltages could reduce our energy-per-bit below \SI{1}{\pico\joule\per\bit}, proving it competitive with the surveyed works.}

\section{Conclusion}

To meet the demands of real-time control algorithms with
increasing performance requirements and the trend toward \revrep{2.5D (chiplet)}{SiP and chiplet integration} technology, we introduce ControlPULPlet, an open-source \revrep{, chiplet-compatible RISC-V controller}{, real-time, heterogeneous multi-core RISC-V controller.} 
It features a \SI{32}{\bit} CV32RT core for fast and deterministic interrupt management, a specialized \gls{dma} for real-time data transfer, and a tightly-coupled \gls{pmca} for compute-intensive \revrep{tasks}{control algorithm acceleration}. 
A flexible AXI4-compatible \gls{d2d} link supports \rev{both} inter-chiplet and on-chip communication. 
Operating at \SI{1.2}{\volt}, \rev{Kairos,} ControlPULPlet's silicon demonstrator \revrep{Kairos}{in 65nm CMOS,} peaks at \SI{290}{\mega\hertz} within \SI{30}{\milli\watt} power envelope during intensive control workloads. 
The \gls{d2d} link enables off-die access at \SI{11.7}{\pico\joule\per\bit} at \SI{1.2}{\volt} IO voltage, with only \SI{1.3}{\pico\joule\per\bit} internal energy, attaining duplex peak transfer rates of \SI{51}{\giga\bit\per\second} at \SI{200}{\mega\hertz}, \SI{2.9}{\percent} and \SI{1.2}{\percent} area and power penalties, and minimal performance degradation on periodic control policies compared to its on-chip counterpart.




\bibliographystyle{IEEEtran}
\bibliography{main}

\newcommand{\missingbio}{has not yet added their bio. This is just a placeholder.}
\newcommand{\lucaphd}[1]{#1 is currently pursuing a Ph.D. degree in the Digital Circuits and Systems group of Prof.\ Benini.}
\newcommand{\ethgrad}[3]{received #1 B.Sc. and M.Sc. degrees in electrical engineering and information technology from ETH Zurich in #2 and #3, respectively.}
\newcommand{\lucagrad}[2]{completed #1 Ph.D. in the Digital Circuits and Systems group of Prof.\ Benini in #2.}
\newcommand{\researchinterests}[1]{research interests include #1.}

\vspace{-0.8cm}
\begin{IEEEbiography}[%
    {\includegraphics[width=1in,height=1.25in,clip,keepaspectratio]%
        {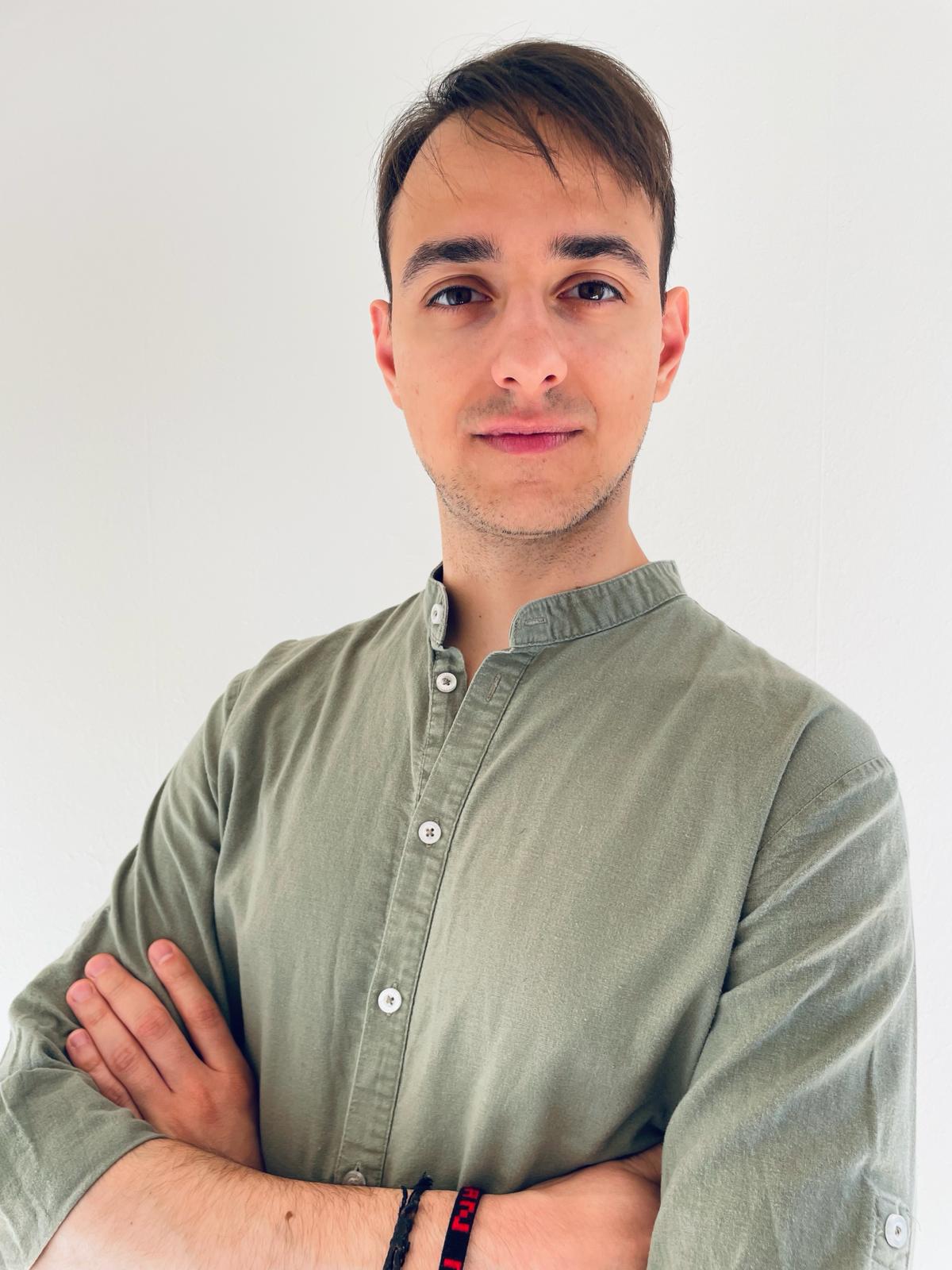}}%
    ]{Alessandro Ottaviano}
    (Graduate Student Member, IEEE)
    received the B.Sc. in Physical Engineering from Politecnico di Torino, Italy, and the M.Sc. in Electrical Engineering as a joint degree between Politecnico di Torino, Grenoble INP-Phelma and EPFL Lausanne, in 2018 and 2020 respectively. %
    \lucaphd{He}
    His
    \researchinterests{real-time and predictable computing systems and energy-efficient processor architecture}
\end{IEEEbiography}
\vspace{-0.8cm}
\begin{IEEEbiography}[%
    {\includegraphics[width=1in,height=1.25in,clip,keepaspectratio]%
        {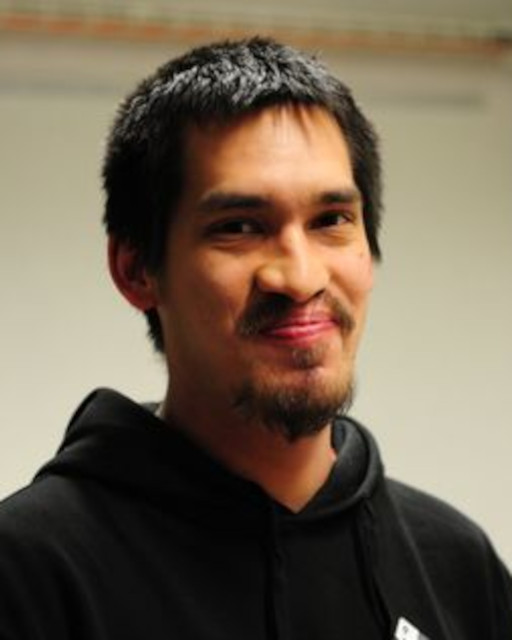}}%
    ]{Robert Balas}
    (Graduate Student Member, IEEE)
    \ethgrad{his}{2015}{2017}
    \lucaphd{He}
    His
    \researchinterests{real-time computing, compilers, and operating-systems}
\end{IEEEbiography}
\vspace{-0.8cm}
\begin{IEEEbiography}[{\includegraphics[width=1in,height=1.25in,clip,keepaspectratio]{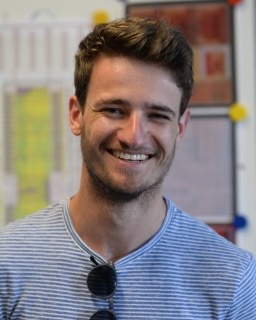}}]
{Tim Fischer} received his BSc and MSc in ``Electrical Engineering and Information Technology'' from the Swiss Federal Institute of Technology Zurich (ETHZ), Switzerland, in 2018 and 2021, respectively. He is currently pursuing a Ph.D. degree at ETH Zurich in the Digital Circuits and Systems group led by Prof. Luca Benini. His research interests include scalable and energy-efficient interconnects for both on-chip and off-chip communication.
\end{IEEEbiography}
\vspace{-0.8cm}
\begin{IEEEbiography}[%
    {\includegraphics[width=1in,height=1.25in,clip,keepaspectratio]%
        {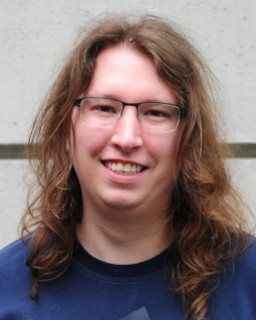}}%
    ]{Thomas Benz}
    (Graduate Student Member, IEEE)
    \ethgrad{his}{2018}{2020}
    \lucaphd{He}
    His
    \researchinterests{energy-efficient high-performance computer architectures, memory interconnects, data movement, and the design of \acrshortpl{asic}}
\end{IEEEbiography}
\vspace{-0.8cm}
\begin{IEEEbiography}[%
    {\includegraphics[width=1in,height=1.25in,clip,keepaspectratio]%
        {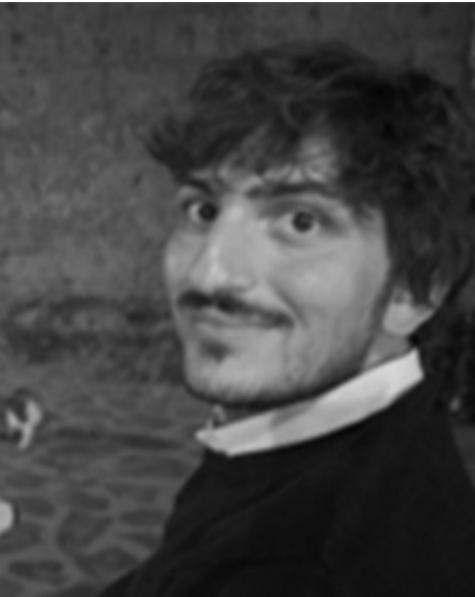}}%
    ]{Andrea Bartolini}
    (Member, IEEE) received the Ph.D. degree from the University of Bologna, Bologna, Italy, in 2013. He is an Associate Professor with the Department of Electrical, Electronic and Information Engineering Guglielmo Marconi, University of Bologna, Bologna, Italy. He has published more than 120 papers in peer-reviewed international journals and conferences and several book chapters with focus on dynamic resource management, ranging from embedded to large-scale HPC systems.
\end{IEEEbiography}
\vspace{-0.8cm}
\begin{IEEEbiography}[%
    {\includegraphics[width=1in,height=1.25in,clip,keepaspectratio]%
        {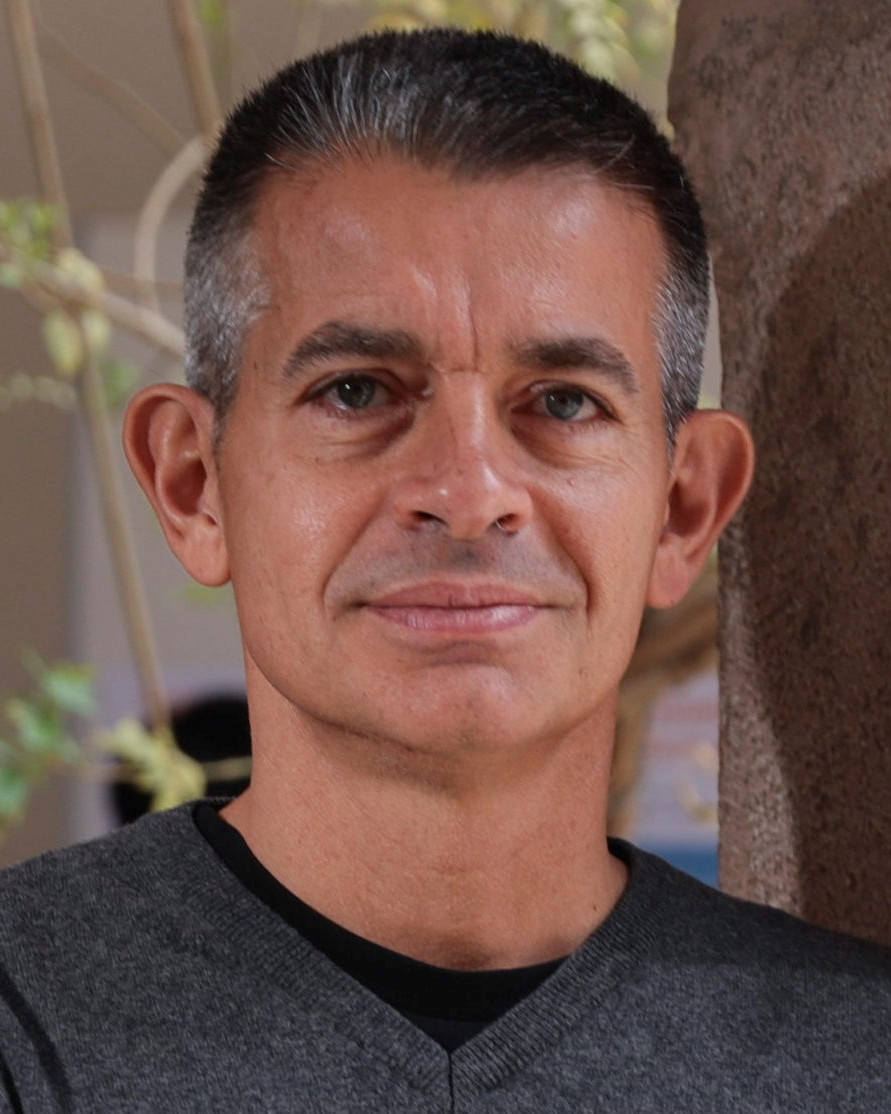}}%
    ]{Luca Benini}
    (Fellow, IEEE) 
    holds the chair of digital Circuits and systems at ETHZ and is Full Professor at the Università di Bologna.
    He received a PhD from Stanford University.
    His research interests are energy-efficient parallel computing systems, smart sensing micro-systems, and machine learning hardware.
    He is a Fellow of the IEEE, of the ACM, a member of the Academia Europaea, and of the Italian Academy of Engineering and Technology.
\end{IEEEbiography}


\end{document}